\documentclass[]{aa}  
\usepackage{natbib}
\bibpunct{(}{)}{;}{a}{}{,} 

\usepackage{graphicx}
\usepackage{txfonts}
\usepackage{hyperref}
\usepackage[svgnames]{xcolor}

\usepackage{aalongtable}

\usepackage{amstext}

\begin{document}

\title{Radial velocity information content of M dwarf spectra in the near-infrared}

\author{P. Figueira\inst{1},
        V. Zh. Adibekyan\inst{1},
        M. Oshagh\inst{1},
        J. J. Neal\inst{1,2},
        B. Rojas-Ayala\inst{1},
        C. Lovis\inst{3},
        C. Melo\inst{4},
        F. Pepe\inst{3},
        N. C. Santos\inst{1,2},
        M. Tsantaki\inst{1}
        }

   \institute{Instituto de Astrof\' isica e Ci\^encias do Espa\c{c}o, Universidade do Porto, CAUP, Rua das Estrelas, PT4150-762 Porto, Portugal\\
     \email{pedro.figueira@astro.up.pt}   
    \and
    Departamento de F\'{i}sica e Astronomia, Faculdade de Ci\^{e}ncias, Universidade do Porto, Portugal
    \and
    Observatoire Astronomique de l'Universit\'{e} de Gen\`{e}ve, 51 Ch. des Maillettes, - Sauverny - CH1290, Versoix, Suisse
    \and 
    European Southern Observatory, Alonso de C\'{o}rdova 3107, Vitacura, Casilla 19001, Santiago 19, Chile 
    }

   \date{}

 
  \abstract
   {}
   {We evaluate the radial velocity (RV) information content and achievable precision on M0-M9 spectra covering the ZYJHK bands. We do so while considering both a perfect atmospheric transmission correction and discarding areas polluted by deep telluric features, as done in previous works.}
   {To simulate the M-dwarf spectra, PHOENIX-ACES model spectra were employed; they were convolved with rotational kernels and instrumental profiles to reproduce stars with a $v.sin{i}$ of 1.0, 5.0, and 10.0\,km/s when observed at resolutions of 60\,000, 80\,000, and 100\,000. We considered the RV precision as calculated on the whole spectra, after discarding strongly polluted areas, and after applying a perfect telluric correction. In the latter option, we took into account the reduction in the number of recorded photons due to a transmittance lower than unity and considered its effect on the noise of the recorded spectra. In our simulations we paid particular attention to the details of the convolution and sampling of the spectra, and we discuss their impact on the final spectra.}
   {Our simulations show that the most important parameter ruling the difference in attainable precision between the considered bands is the spectral type. For M0-M3 stars, the bands that deliver the most precise RV measurements are the Z, Y, and H band, with relative merits depending on the parameters of the simulation. For M6-M9 stars, the bands show a difference in precision that is within a factor of $\sim$2 and does not clearly depend on the band; this difference is reduced to a factor smaller than $\sim$1.5 if we consider a non-rotating star seen at high resolution. We also show that an M6-M9 spectrum will deliver a precision $\text{about}$ two times better as an M0-M3 spectra with the same signal-to-noise ratio. Finally, we note that the details of modeling the Earth atmosphere and interpreting the results have a significant impact on which wavelength regions are discarded when setting a limit threshold at 2-3\%. The resolution element sampling on the observed spectra plays an important role in the atmospheric transmission characterization. As a result of the multiparameter nature of the problem, it is very difficult to precisely quantify the impact of absorption by the telluric lines on the RV precision, but it is an important limiting factor to the achievable RV precision. }{} 
   

  \authorrunning{P. Figueira et al.}

  \maketitle
  \titlerunning{RV information content of M dwarfs in the nIR}
%

\section{Introduction}\label{sec:Intro}

The technique of spectroscopy is central to the study of stars and has allowed astronomy to gather a significant body of knowledge from the few photons a star provides. During the past 20 years, spectroscopy was extensively applied in an emerging field in astronomy: the study of extrasolar planets. Following the discovery of 51\,Peg\, b in 1995 \citep{1995Natur.378..355M}, more than 1900 planets were discovered, with masses and radii down to those of Earth \citep[e.g.,][]{2012Natur.491..207D, 2013Natur.494..452B, 2013Natur.503..377P}. The radial velocity (RV) technique, with which the very first planet around a solar-type star was found, is still one of the most widely used detection methods; it is the main contributor to our knowledge of the mass of known exoplanets. The hunt for the exoplanet with the lowest mass pushed the RV precision down to the subm/s domain and motivated the construction of instruments such as ESPRESSO, which aims at a precision of 10\,cm/s \citep{2012SPIE.8446E..1RM, 2014AN....335....8P}. 

The RV signature of a planet scales with the mass of the star with $M_*^{-2/3}$ and with the planetary orbital period with $P_{orb}^{-1/3}$. For an Earth-mass planet orbiting inside the habitable zone around a solar-type star, the RV amplitude is 10 cm/s, while for a planet with the same characteristics but orbiting an M7 dwarf, the RV amplitude is larger than 1\,m/s. This is due to both a lower host mass and a closer habitable zone (a consequence of the lower luminosity output of these hosts). This relatively high amplitude contributed to an increased interest in the search for exoplanets around the low-mass M-dwarf (0.5-0.08\,M$_{\odot}$) and led to the first estimates of the fraction of M dwarfs hosting Earth-mass planets inside the habitable zone \citep{2013A&A...549A.109B}. The intrinsic faintness of the stars in the visible domain limited these surveys to the brightest one hundred stars of our neighborhood and, along with activity, photon noise contribution to the noise budget proved to be the limiting factor. Spurred by the abundance of exoplanets, the exoplanet hunters did not rest in their efforts, and in the last several years, a new research direction started to gain momentum. Since M dwarfs have their peak emission at near infrared (nIR) wavelengths, and with the advent of high-resolution nIR spectrographs (i.e., with a resolving power $R$ $\equiv$ $\lambda\,/\,\Delta \lambda$ higher than 50\,000, where $\lambda$ denotes wavelength)\footnote{Some authors prefer a different definition, in which the resolving power is defined by the mentioned equation, but is a property of the grating, and the resolution of the spectrograph is dictated by the limited slit width. We prefer avoid pursuing such technical details and use the terms spectral resolution and resolving power interchangeably.}, the knowledge gathered on deriving precise RV started to be transferred to the near-IR (nIR) domain. The first RV studies in the nIR proved that a precision down to the m/s level could be reached \citep[e.g.,][]{2010A&A...511A..55F, 2010ApJ...713..410B}, motivating the development of dedicated nIR spectrographs with the declared goal of detecting Earth-mass exoplanets orbiting M dwarfs.

Several of these spectrographs are becoming a reality. The list of pioneering instruments is headed by SPIRou, the nIR spectropolarimeter for the CFHT \citep{2013sf2a.conf..497D, 2014SPIE.9147E..15A}, the visible-infrared echelle spectrograph CARMENES \citep{2014SPIE.9147E..1FQ}, the Habitable Zone Planet Finder \citep[HZPF,][]{2012SPIE.8446E..1SM}, and the Infrared Doppler instrument for the Subaru telescope \citep[IRD,][]{2014SPIE.9147E..14K}\footnote{To this list the already existing spectrograph GIANO \citep{2014SPIE.9147E..1EO} and future upgraded version of CRIRES, CRIRES+ \citep{2014SPIE.9147E..19F} need to be added; without being designed to specifically achieve the mentioned goal, these spectrographs will certainly be used to perform RV measurements.}. The common denominator for these spectrographs is the choice of high resolution and high stability, so that RV can be measured with high precision. High spectral resolution is required for spectral lines to be resolved and to determine their center with precision; since the RV precision is known to be proportional to the resolution to the power of 3/2 \citep[e.g.,][]{1992ESOC...40..275H, 2014Natur.513..358P}, the current designs aim at R\,$>$\,50\,000. For slowly rotating stars, a resolution of 100 000 might prove advantageous (as chosen for visible spectrographs), but the large number of fast-rotating M dwarfs suggests \citep[e.g.,][]{2012AJ....143...93R} that for most stars a very high resolution will not lead to an appreciable increase in precision (as seen in the very same work). The high stability requirement is in part built-in because nIR spectrographs have to operate in temperature-controlled environments to reduce thermal background emission and its fluctuations. The stability is boosted by operating them in vacuum and using homogeneous lighting feed systems such as octagonal fibers or mechanical scrambling \cite[e.g.,][]{2014SPIE.9147E..6BR} in addition to improved thermal stability.

The main difference among these spectrographs is the wavelength domain they cover. SPIRou will operate in the YJHK bands, CARMENES in the visible and YJH bands, HZPF in the ZYH, and IRD in the YJH. Other than the difference in construction, operating complexity, and associated cost, these design choices were made based on studies on the RV information content of the nIR bands. \cite{2010ApJ...710..432R} used model spectra calculated with the PHOENIX code to compare the achievable precision in nIR bands YJH with that obtained in the optical. The authors considered spectral types of M3, M6, and M9 and compared the information in each of the wavelength bands with that of the visible V band. The authors discussed the impact of observing at different resolutions and used simple scaling laws to consider the effect of an imperfect wavelength calibration, starting from measured precision values for ThAr lamps \citep{2007A&A...468.1115L, 2008ApJS..178..374K} and gas cells \citep{1996PASP..108..500B}. The work of \cite{2010ApJ...710..432R} was later on complemented by that of \cite{2011A&A...532A..31R}, who extended the analysis to ultra-cool M dwarfs and L-dwarfs; very importantly, the K band was also considered, unlike in previous works. In both studies the authors chose to completely mask the spectral lines that fall within $\pm$30\,km/s (the maximum RV amplitude of the Earth around the Sun) of atmospheric transmission features deeper than 2-3\%. This procedure removes a significant fraction of the nIR bands from the analysis, especially at longer wavelengths, because of the stronger absorption from molecules such as water and methane. While a good first approximation, this approach is probably too conservative. Recent developments in modeling the Earth's atmosphere will allow for at least a fraction of the information to be recovered. The developments in this direction in the past few years have been remarkable: from \cite{2010A&A...524A..11S} to the recent web interface TAPAS \citep{2014A&A...564A..46B}, the {\it Telfit} \citep{2014AJ....148...53G} and {\it Molecfit} codes \citep{2015A&A...576A..77S}, to the recent PCA-based approach of \cite{2014SPIE.9149E..05A}, we have come a long way in the recent years. We are even now starting to analyze the effect of small amplitude or unresolved stellar lines on stellar visible spectra \citep{2014A&A...568A..35C}.

It has long been known that nIR spectral atlases are incomplete; they identify and reproduce only a fraction of existing transitions with the correct strength even for the most abundant atoms and molecules \citep{2007A&A...473..245R, 2010ApJS..186...63R}. As such, the usage of the most recent stellar atmospheric models is necessary to reproduce the stellar features in the best way possible.

In this work we extend and improve upon the previous works by considering stars with effective temperatures similar to those of M0 to M9 and evaluating the information content in the ZYJHK bands. We consider several cases of atmospheric transmission correction, from the overoptimistic perfect correction to the more restrictive conditions of the previous authors. Very importantly, we use the most recent models that allow us to achieve the best characterization possible of the stellar spectra. In Sect.\,2 we describe the simulations performed, and in Sect.\,3 we present our results and interpret them. In Sect.\,4 we discuss the implications of our findings and conclude in Sect.\,5.


\section{Models and simulations}\label{simuls}

\subsection{Models}

In our analysis we used model spectra calculated with the PHOENIX code \citep[e.g.,][]{1999ApJ...512..377H, 2001ApJ...556..357A} for effective temperatures of 3900, 3500, 2800, and 2600\,K that correspond to stellar types of M0, M3, M6, and M9. As template spectra we selected those delivered by the most recent PHOENIX-ACES models described in \cite{2013A&A...553A...6H}. The surface gravity value chosen was 4.5, and for the metallicity we used the solar value. These values were chosen because they are common in field M-dwarfs. To convert the flux (energy per wavelength bin) into photon counts (photons per wavelength bin), we divided the flux by the energy of a photon. Neglecting the multiplicative constants, this corresponds to multiplying the flux values by the respective wavelength. We then convolved the spectrum with a rotational profile following the description presented in \cite{2008oasp.book.....G} using a linear limb-darkening coefficient of $\epsilon$\,=\,0.6. For projected stellar rotational velocity $v.\sin{i}$ values we considered the cases of 1.0, 5.0, and 10.0\,km/s. We considered these values as representative of slowly rotating standard and fast-rotating M dwarfs (for RV measurement purposes) of the solar neighborhood, following \cite{2012AJ....143...93R}. 

We then convolved the rotation-broadened spectra with Gaussian instrumental profiles (IP) with FWHM equivalent to spectral resolution values of 60\,000, 80\,000, and 100\,000. It is important to note that the resolution of the PHOENIX-ACES models is 500\,000 for the wavelength domain of interest to us, while when we convolve
the spectra, we assume that the spectrum has an infinite resolution (i.e., it was not broadened by any mechanism other than those involved in line formation itself, such as thermal excitement). When convolving a spectrum with 500 000 resolution with a Gaussian with FWHM representative of an IP of a spectrograph with R\,=\,100 000 (of  2.0$\times$10$^{-5}\,\mu$m at 1\,$\mu m$), the final unresolved line will have a FWHM less than 2\% larger than if convolved with a spectrum of infinite resolution, as assumed. Therefore we consider that the error introduced by this convolution is negligible, even for the case of a non-rotating star at the highest resolution considered. 

When performing a convolution with a profile, we assume that the profile has unitary area, and as such, the convolution is an area-preserving operation. However, for variable $\Delta \lambda$ steps, the different discretization of the profile as we perform the convolution as a function of wavelength leads to a multiplication by a function with a slightly different area. This might lead to a noticeable effect on the results for very different sampling and/or high flux values, two cases for which the approximation of constancy of area might break. Unfortunately, this is the case for PHOENIX-ACES models, just like for many other models. \cite{2013A&A...553A...6H} stated that for the nIR domain covered here the resolution of the profile is approximately constant, and in fact this is obtained by using a $\Delta \lambda$ step that depends on wavelength and is increased every 500 nm. This change in step leads to appreciable changes in the convolution kernel area, which in its turns translates into appreciable flux differences as we move from a domain with a value of $\Delta \lambda$ to another with a different value. To correct for this effect, the resulting flux value needs to
be normalized by the convolution kernel area (which, theoretically, should be of unity). To implement this, we normalized the result of the previously described convolution by the one obtained by performing a convolution of a spectrum of constant flux value of 1.0 (this convolution should by definition yield a result of 1.0). In a nutshell, the result of our convolution procedure is the PHOENIX-ACES spectrum convolved with a rotational kernel and a Gaussian representing the IP, divided by a unitary function with the same wavelength grid and convolved by the same Gaussian function and IP. 

When simulating stellar spectra collected by a high-resolution spectrograph, it is important to correctly reproduce the sampling
effect. The sampling of a spectrum is the number of recorded pixels per resolution element. For echelle spectrographs, for which $\Delta\lambda$\,/\,$\lambda$ is constant, the number of pixels per resolution element is also constant. The Nyquist theorem tells us that at least two pixels are required to sample the resolution element for a significant fraction of the information not to be lost, and the resolution degraded. However, modern spectrographs are usually designed to record spectra with a higher sampling value, and HARPS, for instance, has a sampling of 3.3 pixels per resolution element \citep{2003Msngr.114...20M}. For our analysis it is therefore important to consider the same sampling for all the spectra, so that the spectra are compared on an even footing; the assumption of a constant value is more important than the value itself, as long as it is larger than 2. For our study, we chose to use the commonly used sampling value of three pixels per resolution element. To reproduce the sampling of the spectra correctly, we considered the starting $\lambda$ value for each of the PHOENIX spectra and interpolated it at intervals given by $\lambda / (3\,R)$.

The bands and wavelengths considered for this study are Z\,(0.83-0.93$\mu$m), Y\,(1.0-1.1$\mu$m), J\,(1.17-1.33\,$\mu$m), H\,(1.5-1.75\,$\mu$m), and K\,(2.07-2.35\,$\mu$m). 

\subsection{Calculating the RV information content of a spectrum and associated precision}

To calculate the theoretical RV precision delivered by a spectrum, we followed the procedure described in \cite{2001A&A...374..733B}. The authors considered high-signal-to-noise spectra and evaluated radial velocity variations, which are very small when compared with the width of the spectral lines\footnote{We recall that this condition is fulfilled if we consider radial velocities at m/s level, as measured from high-resolution spectra, since a FWHM is always larger than 2\,km/s, as imposed by the IP contribution alone.}. From their Eq.\,(11) and (12) it follows that the RV uncertainty associated with the information content of a given spectra is given by

\begin{equation}\label{RV_rms}
 RV_{rms} = \frac{c}{Q\sqrt{N_e}} = \frac{c}{\sqrt{\sum_{i} W(i)}}
,\end{equation}

in which $c$ is the speed of light in vacuum, $Q$ the quality factor of a spectrum, and $N_e$ the total number of photoelectrons collected inside the wavelength range of interest. However, the precision can only be calculated using the concept of optimal pixel weight $W(i)$ for each of the pixels $i$ that compose the spectra,

\begin{equation}\label{weight_equation}
W(i) = \frac{\lambda^2(i)(\partial A_0(i)/\partial \lambda(i))^2 }{ A_0(i) + \sigma_D^2}
,\end{equation}

in which $\lambda(i)$ and $A_0(i)$ are the values of each pixel wavelength and flux, respectively. The weight will be proportional to the information content of the spectrum, given by the derivative of the amplitude, and calculated following \cite{1985Ap&SS.110..211C}. The denominator of the previous equation is the variance of the flux of the pixel $A_0(i)$, depending on the flux value itself and on the detector noise $\sigma_D$. In this paper we exclusively
consider the high signal-to-noise ratio regime, so we can approximate $A_0(i) + \sigma_D^2 \sim A_0(i)$. When considering the average RV as delivered by several slices of a spectrum, the error on the average is given by the error on a weighted average

\begin{equation}\label{weighted_RV}
 \overline{RV_{rms}} = \frac{1}{\sqrt{\sum_{i} (\frac{1}{RV_{rms}(i)})^2}} .
\end{equation}

To take into account the pollution introduced by the telluric lines in the spectrum, previous works have masked out areas that were affected by deep telluric absorption and discarded them when calculating the precision attributable to a given band. This implies discarding all the stellar lines that are polluted by the telluric lines, which have a relative motion from -30 to +30\,km/s due to the relative motion of the Earth to the fixed star (whose systemic RV we assume to be 0). While a reasonable assumption, we consider this to be an extreme-case scenario, and it leads to discarding a significant fraction of the information content on a spectrum. Several teams have sucessfully reproduced the detailed features of atmospheric transmission spectrum, and it can be assumed that these allow us to recover a fraction of RV information content associated with the stellar features polluted by these lines. Following this rationale, we apply Eq.\,\ref{RV_rms} and \ref{weight_equation} under three different conditions:

\begin{enumerate}
 \item Over the whole spectral band -- {\it \textup{non-polluted spectra, inaccessible to us, as if the observations were obtained from space}}.
 \item Discarding pixels at less than 30\,km/s from telluric absorption features deeper than 2\% -- {\it \textup{fraction of the spectra not polluted by telluric lines, the same as used by \cite{2010ApJ...710..432R} and \cite{2011A&A...532A..31R}}}. \item Considering that the variance represented in the denominator, that is, the photon noise contribution to the spectrum, is amplified by the telluric correction, and is given by the measured photon count $A_{0}(i)$ divided by the telluric transmission at that wavelength squared $T(\lambda(i))^2$-- {\it \textup{total spectral information after a perfect correction is performed}.}
\end{enumerate}

The first condition is only presented for comparison; it can never be experienced from a ground observatory, on which high-resolution spectrographs are currently installed. On the other hand, the last condition is extremely important and merits a detailed explanation. Even if we consider the best-case scenario in which our observations and models are of very high quality and the atmospheric transmission is corrected for in a perfect way, the reduced transmission will have an impact on the measured flux. When we divide the measured flux $A_{0}(i)$ by the transmission $T(\lambda(i))$ to obtain the corrected spectrum $A_{0}'(i)$, we are actually dividing the scatter by the same value. As a consequence, the variance term in the denominator of Eq.\,\ref{weight_equation} becomes $A_{0}'(i)$ / $T(\lambda(i))^2$. 

To perform the analysis of the two last points, we used the software TAPAS \citep{2014A&A...564A..46B} to simulate the telluric spectrum. We modeled the atmospheric transmission as seen from the La Silla Observatory at an airmass of 1.2 ($z$\,=\,33.5$^o$). This is the standard value as considered by several exposure time calculators and is in line with the average airmass as measured in our HARPS programs dedicated to high-precision RV measurements. We considered atmospheric models as delivered by the embedded ARLETTY interface and recovered one spectra per week for 2014 as representative of the long-term variations of absorption over the year. When performing the query for the model, we requested a resolution of R\,=\,100 000 (the highest considered in this work) and a sampling factor of 10. Since our stellar spectra have a sampling of 3.3, one pixel of our spectra corresponds, on average, to approximately three pixels in our telluric model.

The precision delivered by these simulations and calculation is a relative precision, and its final value will depend on the S/N of the spectra. We are only interested in relative precision, and therefore we normalized the spectra to deliver a S/N per resolution element (i.e., in the three pixels which we assumed as used for sampling) of 100 as observed at the center of the J band, at 1.25 microns (in an area without any detectable lines). 

\section{Results and interpretation}

\subsection{Results}

\begin{figure*}
\centering
\includegraphics[width=16cm]{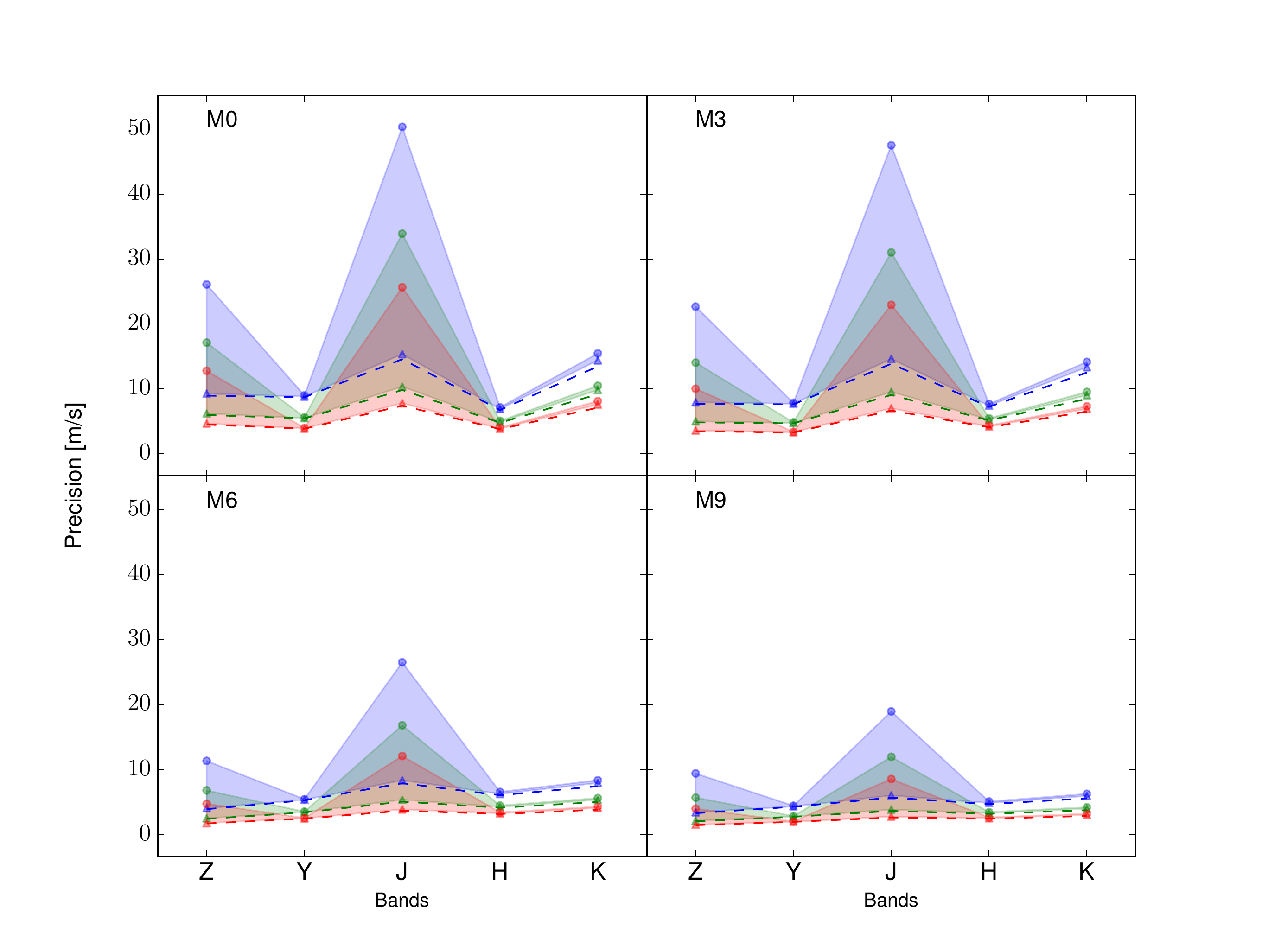}

\caption{Precision achieved as a function of spectral band for stars with a  rotational velocity of $v.\sin{i}$\,=\,1.0\,km/s and spectral types M0, M3, M6, and M9. The dashed line represents the theoretical limits imposed by condition 1, and the filled area represents the values within the limits set by conditions 2 ({\it circles}) and 3 ({\it triangles}); blue, green, and red represent the results obtained for resolutions of 60\,000, 80\,000, and 100\,000, respectively. The spectra were normalized to have a S/N of 100 per resolution element as measured at the center of the J band (see Sect.\,\ref{simuls} for details).}\label{Res1}

\end{figure*}

\begin{figure*}
\centering
\includegraphics[width=16cm]{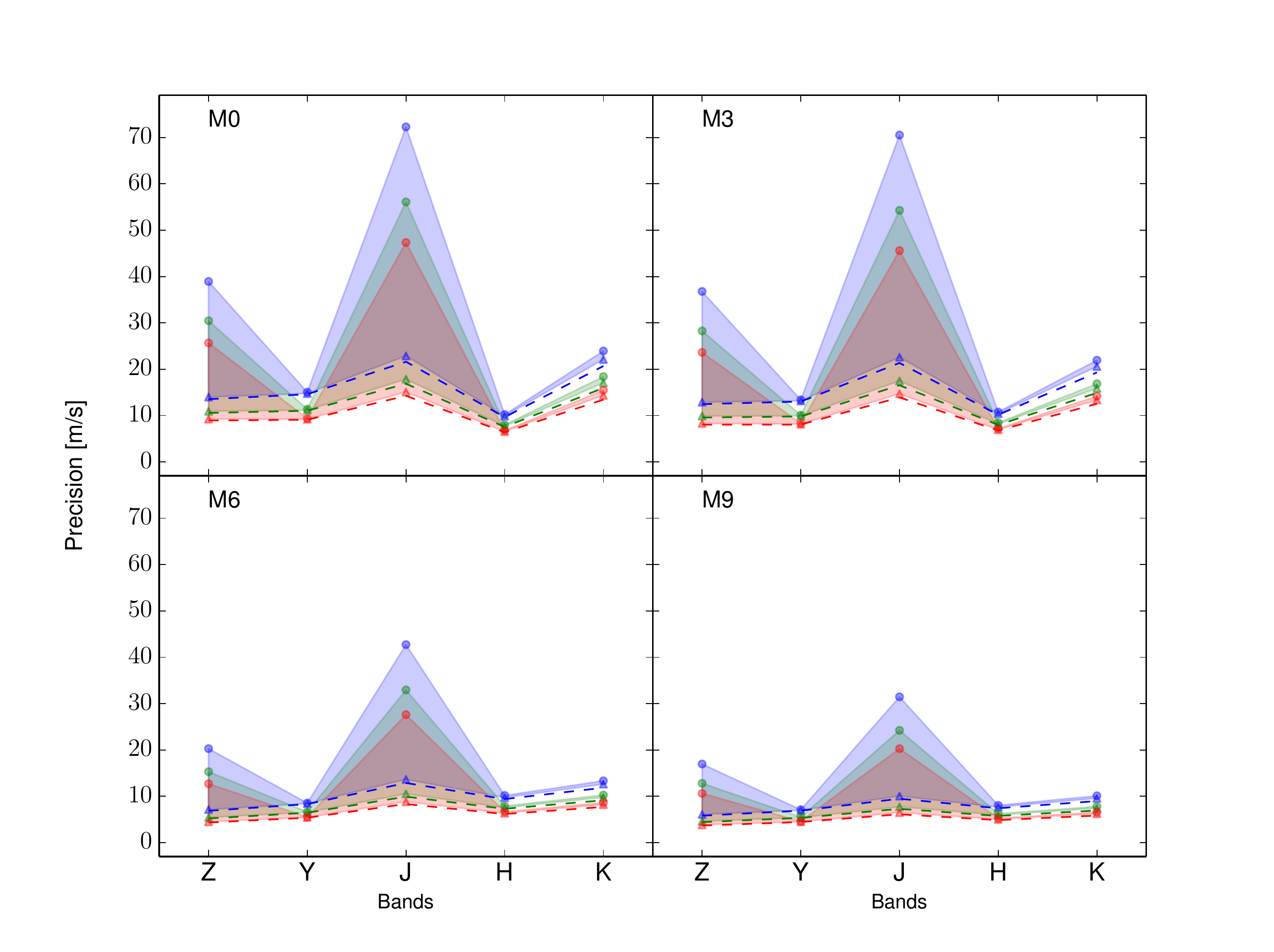}

\caption{Precision achieved as a function of spectral band for stars with a rotational velocity of $v.\sin{i}$\,=\,5.0\,km/s and spectral types M0, M3, M6, and M9. The dashed line represents the theoretical limits imposed by condition 1, and the filled area represents the values within the limits set by conditions 2 ({\it circles}) and 3 ({\it triangles}); blue, green and red represent the results obtained for resolutions of 60\,000, 80\,000, and 100\,000, respectively. The spectra were normalized to have a S/N of 100 per resolution element as measured at the center of the J band (see Sect.\,\ref{simuls} for details).}\label{Res2}

\end{figure*}

\begin{figure*}
\centering
\includegraphics[width=16cm]{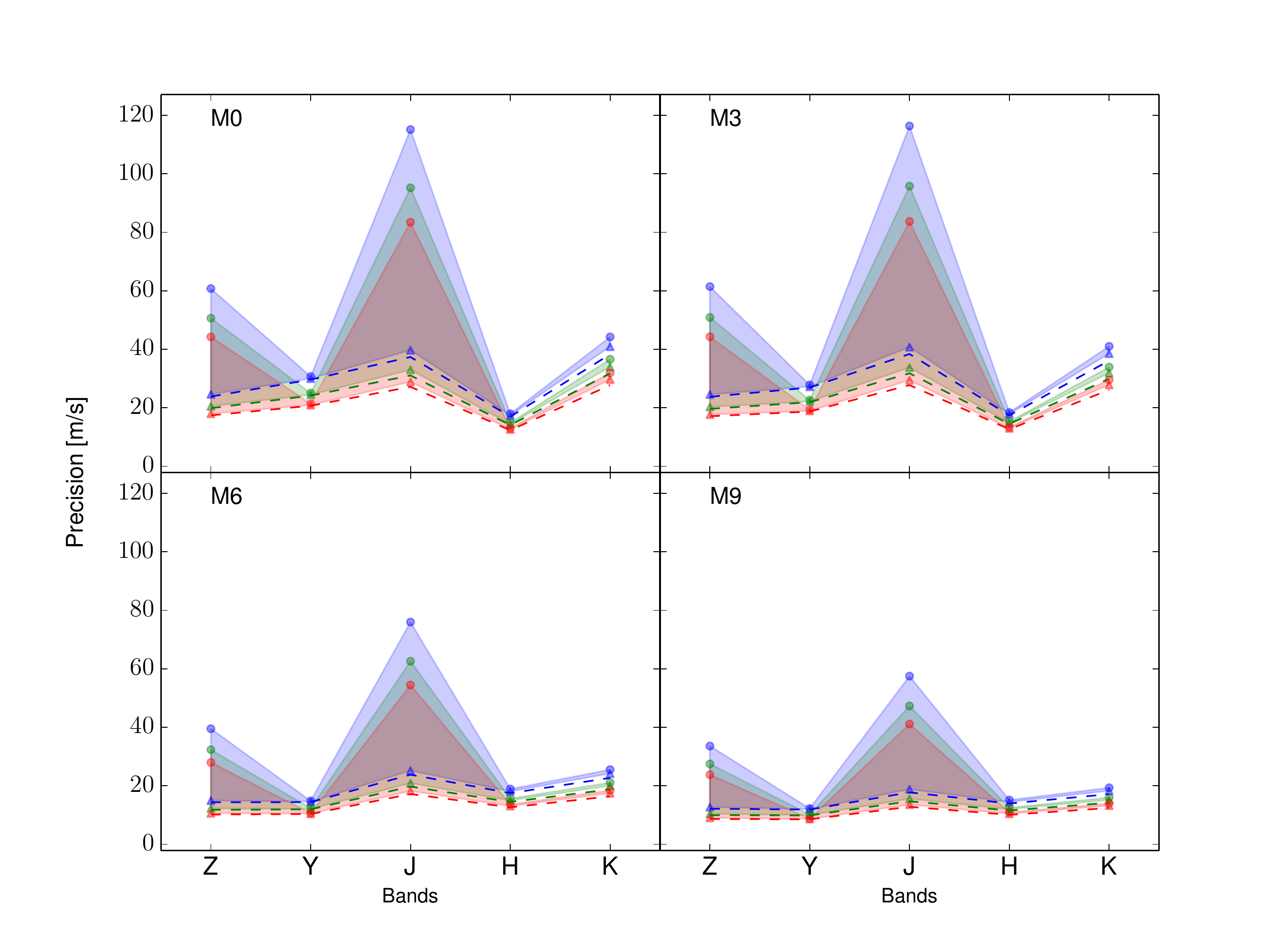}

\caption{Precision achieved as a function of spectral band for stars with a rotational velocity of $v.\sin{i}$\,=\,10.0\,km/s and spectral types M0, M3, M6, and M9. The dashed line represents the theoretical limits imposed by condition 1, and the filled area represents the values within the limits set by conditions 2 ({\it circles}) and 3 ({\it triangles}); blue, green and red represent the results obtained for resolutions of 60\,000, 80\,000, and 100\,000, respectively. The spectra were normalized to have a S/N of 100 per resolution element as measured at the center of the J band (see Sect.\,\ref{simuls} for details).}\label{Res3}

\end{figure*}

The RV precision as estimated from conditions 1, 2, and 3 is presented in Table\,\ref{TableData} and plotted in Figs.\,\ref{Res1}, \ref{Res2}, and \ref{Res3}, corresponding to stars with $v.\sin{i}$ values of 1.0, 5.0, and 10.0\,km/s, respectively. The first point to note is that, as expected, the highest precision is obtained for the lowest rotational velocity (Fig.\,\ref{Res1}), which corresponds basically to a non-rotating star, that is, to a star on which measured broadening is dominated by the IP. Also unsurprisingly, the highest resolution spectra permit the highest precision RV measurements. As we move toward faster rotating stars, the precision decreases and the gain in increasing the resolution decreases as well because the rotational broadening dominates other broadening mechanisms. This general behavior has been discussed in previous studies and can be derived from rules-of-thumb such as those presented in \cite{1992ESOC...40..275H}.

The comparison between the precision achievable in the different bands and the objective of this paper is much less intuitive. The first point to note is that under the same conditions of R and $v.\sin{i}$, the precision as measured on a given spectral type never varies by a factor larger than $\sim$3 from one band to the other. The results for the different rotational velocities and spectral types can be outlined as follows: 

\begin{enumerate}

\item {\it slowly rotating M0 dwarfs:} when observed at a resolution of 100 000, the spectral bands that yield the highest RV precision are the Y and H bands, which have similar precision; these are followed by the Z band. If we reduce the resolution to 60 000, the H band becomes the most precise, followed by Y and Z, delivering similar values. 

\item {\it slowly rotating M3 dwarfs:} the picture is similar, but now Y and Z deliver the highest precision, with H following; the gap between these three bands is reduced to a negligible level if we operate at a resolution of 60 000. 

\item {\it slowly rotating M6 and M9 dwarfs:} the dependence of the precision on bands is smoothed and becomes very similar, with a variation lower than a factor of $\sim$1.5 between the lowest and highest precision bands if we neglect telluric pollution or correction effects. 

\item {\it faster rotating M0 and M3 stars (5 and 10\,km/s):} the H band now allows the most precise measurements, with the Z and the Y bands following. 

\item {\it faster rotating M6 and M9 stars:} the dependence of the precision on the bands is very similar to the analogous case for non-rotating stars, but the difference between bands is slightly more pronounced. Z and Y bands now deliver slightly more precise measurements, followed by the H band. 

\end{enumerate}
 
When comparing different spectral types (or effective temperatures), and for the spectra considered, M6 and M9 provide a precision a factor of $\sim$2 better than their hotter counterparts. It is also interesting to note that the gain in precision when moving from a resolution of 60 000 to 80 000 is higher than from 80 000 to 100 000. 

\subsection{Interpretation}

\begin{figure*}
\centering
\includegraphics[width=14cm]{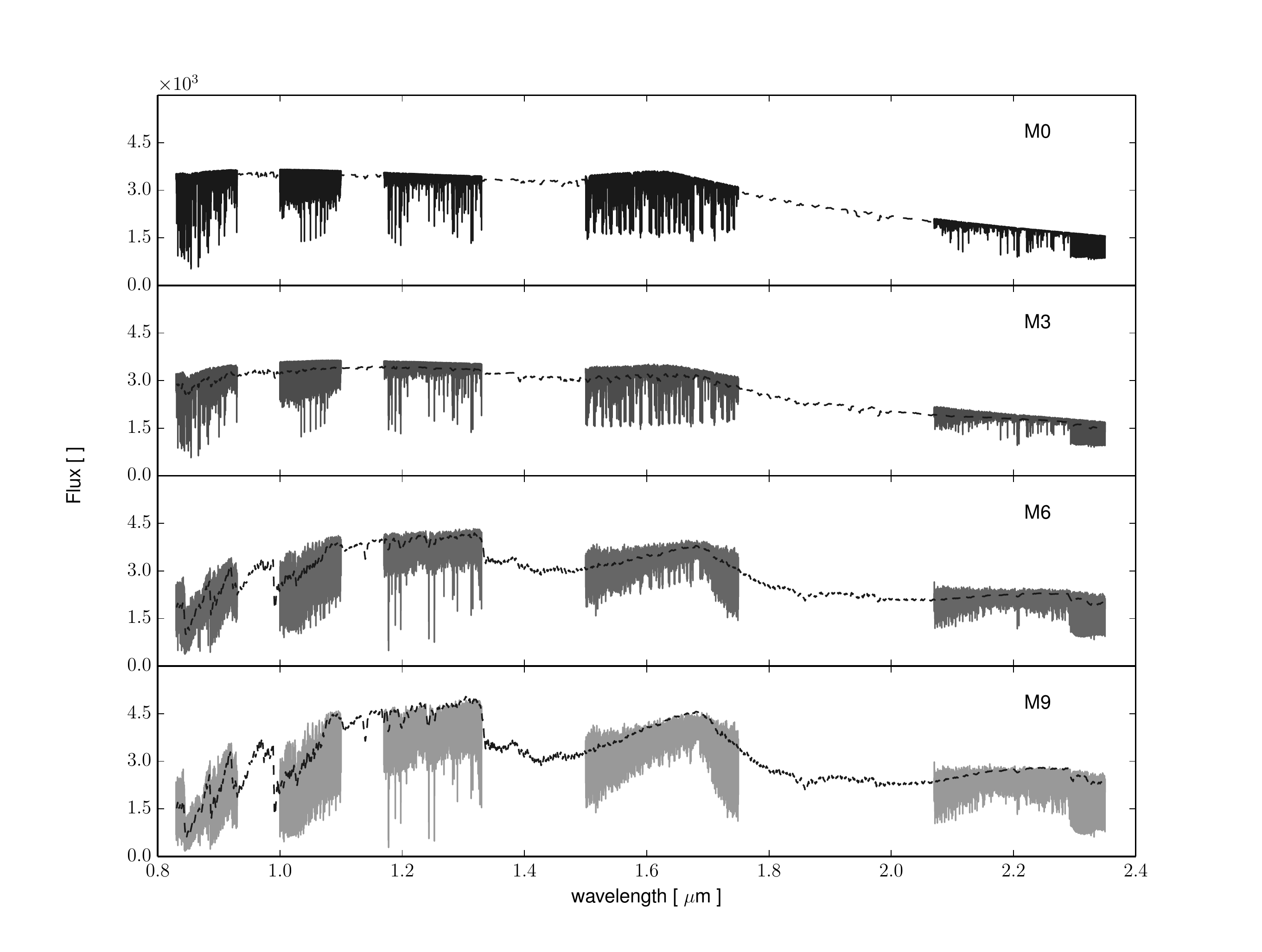}

\caption{Flux per pixel as a function of wavelength for the $v.sin{i}$\,=\,1.0\,km/s and spectral types M0, M3, M6, and M9 ({\it top} to {\it bottom} panels) spectra at a resolution of 100\,000. For comparison, the offset original spectra is presented after smoothing with a boxcar of 2000 pxl ({\it dashed line}), a crude approximation of the stellar continuum. Flux units are arbitrary.}\label{Mdwarfs_flux}

\end{figure*}

To better understand these results, we consider the M-dwarf spectra as recorded in the different bands. The flux per pixel on the ZYJHK bands and for the R\,=\,100\,000 and $v.\sin{i}\,=\,1$\,km/s is presented in Fig.\,\ref{Mdwarfs_flux}, where we overplot the PHOENIX-ACES spectrum after boxcar smoothing for comparison (approximately representative of what the pseudo-continuum of the spectrum would look like). The flux per pixel does not change significantly across the considered wavelength domain
for any spectral type, and it is always within a factor of 3 of that of our reference, the center of the J band. This corresponds to a S/N difference of $\sim$1.7, which translates directly into the achievable precision, all other properties remaining constant \citep[e.g.,][]{1992ESOC...40..275H}. 

From Fig.\,\ref{Mdwarfs_flux} we can perceive a clear difference between the M0-M3 and M6-M9 stars. For the M0-M3 stars the Z and Y flux is essentially the same as that of J band, and Z band contains the deepest lines. There are no strong continuum-shaping water band absorption features at the two edges of H band, and the lines are still separated spectroscopically at high-resolution. Toward M6-M9, the flux in the Z band decreases appreciably with decreasing temperature, and the flux in the Y and the H bands decreases slightly, while that in the K band increases. The well-defined spectroscopic lines are also transformed into bands, with the water absorption features on the two sides of the H band clearly visible. A close-up of the H- and K-band spectra, shown in Fig.\,\ref{Hband_flux} and Fig.\,\ref{Kband_flux}, shows that toward later-type stars, the number of features increases significantly. The presence of these same features contributes to a much higher information content as evaluated by Eqs.\,\ref{RV_rms} - \ref{weighted_RV}, and this explains the increased precision on lower-mass M dwarfs as well as the increased precision in the K band when compared with the earlier-type stars. The spectra also become remarkably more homogeneous in wavelength, with the contribution from specific features like the CO band-heads being diluted among the contribution from the large number of features we can densely record over the whole nIR range. For reference, we plot the stellar and telluric spectra we used are provided in Appendix\,\ref{AppendixPlot}.

\subsection{Impact of telluric correction and flagging}\label{AtmFlag}

It is important to have an estimate of the gain in RV precision by performing a correct telluric absorption when compared to discarding the polluted wavelength sections. The bands considered contain absorption features created by molecules with different properties, which can be corrected for at different precision levels. For instance, the numerous and sparsely located H$_2$O features, which are highly dependent on atmospheric conditions, are much more difficult to correct for than those associated with O$_2$, which is mixed throughout the atmosphere up to an altitude of 80\,km. The effort required to correct for atmospheric
features in a precise way can therefore be very different and should be weighted against the RV precision gain achievable through such a correction. 

A first point to address is the difference between the theoretical limit derived by ignoring the atmosphere contribution and the one obtained considering a perfect correction. The previous result shows that the difference is indeed small. This is expected because the average and median transmission inside the discussed bands is indeed high, in spite of the numerous telluric absorption features. For the Z band the mean transmission is 0.96, for the Y band 0.99, for the J band 0.95, for the H band 0.97, and for the K band 0.95. The median is 0.98 for the Z and K band and 0.99 for the others. From Eqs.\,\ref{RV_rms} and \ref{weight_equation} we can deduce that if we consider an overall transmission of T throughout the spectrum, this leads to a decrease in precision of $RV_{RMS}'$\,=\,$RV_{RMS}/T$. This shows that if we succeed at perfectly correcting the spectra, the loss in precision triggered by the loss in flux is expected to be on the order of a few percent. This highlights the fact that for spectroscopy one could and should extend the data collection and analysis to beyond these historically photometrically motivated bands, and it would still
not be strongly affected by telluric pollution.  

Another important point is to quantitatively assess the gain associated with a perfect atmospheric absorption correction versus discarding the polluted areas. Figures\,\ref{Res1}, \ref{Res2}, and \ref{Res3} show that the most affected bands are the Z and $H$ bands. The deep atmospheric features in these two bands can certainly explain the reduction in precision by a factor of two; however, the absence of an appreciable reduction in precision in the likewise strongly polluted K band casts some doubts on the result.

To discard a part of the spectrum, it is assumed that a telluric spectrum with the same resolution is available and perfectly
reproduces the observing conditions. Central to this issue is the sampling of the atmospheric spectrum. As noted before, when creating our atmospheric spectra using TAPAS, we requested a sampling of the R\,=\,100 000 Gaussian profile of 10. For all features composed of at least three consecutive pixels with a depth smaller than the threshold of 2\%, we removed the spectral domain within reach of a $\pm$30\,km/s velocity shift in our analysis. The number and position of the discarded pixels can
vary strongly as a result of the many features with a depth around 2\% and depending on the criteria considered for the line sampling and its detection. 

To illustrate this aspect, we consider the situations in which we discard wavelength values for which we require 2, 3, 4, or 5 consecutive pixels flagged as below the transmission threshold in our atmospheric model (at a $\Delta \lambda$ smaller than the one induced by a RV displacement of 30\,km/s). Since the telluric spectrum resolution element was sampled by 10 points, the requirement of 2, 3, 4 or 5 consecutive pixels corresponds, in practice, to sampling the spectral line as observed by a  spectrograph with (at least) 5, 3.3, 2.5 or 2 pixels in order to detect an appreciable absorption. The smaller the number of pixels per resolution element, the lower our ability to identify atmospheric absorption features\footnote{For simplicity we assumed that we observe the stellar spectra and characterize the atmospheric spectra with the same spectrograph. Advanced set-ups can be considered, in which this is not the case.}. It is interesting to evaluate the different flagging conditions, however, because a well-benchmarked model will provide information with both a sampling and resolution superior to those of our spectrograph. Overall, different conditions lead to different discarded areas: for a requirement of 5, 4, 3, and 2 consecutively flagged pixels in the vicinity of a wavelength in order to discard it, we have a remaining $\sim$100, 98, 87 and 53\,\% non-discarded area in the bands considered in this work. This result in itself
underlines a seldom considered advantage in opting for a sampling level above the Nyquist sampling. This result is clearly highly dependent on the model chosen, but it illustrates very well that a large area of the near-infrared is characterized by an atmosperic transmission very close to the threshold value of 98\%. We present the result of our calculus in Figs.\ref{Res_4pixel} and \ref{Res_2pixel}, which are the equivalent of Fig.\,\ref{Res1}, but for the 4 and 2 consecutive pixel flagging conditions (the 5-pixel condition is not presented because it is very similar to the 4-pixel condition). They show remarkably different results on the achievable precision when applying condition 2 because very different areas are discarded.

Another aspect that affects the same calculus is the systemic RV of the observed star. We assume the M dwarf spectra to be at rest relative to us, which is seldom the case. However, most stars observed in extrasolar planet searches have an average RV of up to 100 km/s, some even more \citep[e.g.,][]{2014ApJ...789..154D}. A large average radial velocity of the star will offset the lines in wavelength when compared to the spectra analyzed here and lead to different pollution scenarios. To test for this effect, we shifted the atmospheric spectra by $\pm$10km\,/s, emulating a reciprocal RV movement by the star. We repeated the computation described previously and present the equivalent of Fig.\,\ref{Res1}, but for an offset of -\,10\,km/s, in Fig.\,\ref{Res_vel-10}, and of +\,10\,km/s, in Fig.\,\ref{Res_vel+10}. By varying the stellar shift applied to the spectra, we can greatly vary the way the different wavelength domains are flagged and the finally achieved precision as reported by condition 2. The two figures
show that the differences can vary greatly depending on the applied shift. For the examples considered, the variation is particularly large in the J band, probably because most of the RV information is localized in a small number of sharp features. When these features overlap with telluric spectra and are thus discarded, the total precision achievable is greatly affected.

In summary, because the flagging of a wavelength region depends on and varies as a function of 1) the sampling of the atmospheric spectra and 2) the RV of the observed star, we advise against using condition 2, or similar, as an indicator of the achievable RV precision.

\section{Discussion}

\subsection{Assumptions and limitations of the simulations}

Throughout this paper we considered that the RV precision was limited by the information content present in the stellar spectra; in doing so, we assumed that the wavelength calibration introduced a negligible error for the whole wavelength domain considered. While this is a reasonable working hypothesis, reality clearly is far more complex. As an example, different spectrographs employ different wavelength calibration procedures with potentially different associated precision values. These were motivated chiefly by the capabilities of different techniques at different wavelengths, but also by the associated costs. For instance, the HZPF will use a laser frequency comb backed by U-Ne lamps \citep{2014SPIE.9147E..7ZH, 2014PASP..126..445H} while CARMENES will opt for U-Ne lamps exclusively \citep{2014SPIE.9147E..54S} and SPIRou for a Fabry-Perot etalon. The japanese IRD will use a laser frequency comb approach exclusively \citep{2014SPIE.9147E..14K}. The development of precise wavelength calibrators is an active line of research, and significant improvement has been made over the past few years \cite[e.g.,][]{2012SPIE.8446E..8EW, 2014A&A...569A..77R}. 

We also assumed that the spectrograph and detector properties are constant enough over the considered wavelength range to have a negligible impact on the achievable RV precision. In particular, detector properties can change significantly from the Y to the K band. Shortward of 1.0\,$\mu$m\footnote{Even though the quantum efficiency cutoff is at 1.2\,$\mu$m, detectors often struggle with extremely low quantum efficiency longward of 1.0\,$\mu$m.}, silicon-based architecture CCDs are preferable because they are characterized by lower readout noise, a more spatially homogeneous electronic response and signature, and because they have been deeply studied by a large community. Longward of the silicon valence gap, a composite metal oxide semiconductor (CMOS) architecture  or HgCdTe detector with CMOS multiplexer is often chosen, which allows for more flexible reading patterns, non-destructive readout and on-chip analysis, at the cost of introducing a more instrument- and pixel-dependent signature. Our work is predicated on the assumption that the impact of a detector across the wavelength range is negligible, and such a condition must be watched carefully. 

It is crucial to understand down to which precision atmospheric lines can be corrected. \cite{2007PASP..119..228B} demonstrated conclusively that using telluric models provides a better correction than using telluric standards, and \cite{2010A&A...524A..11S} and \cite{2010ApJ...713..410B} reached a residual level of around 1\%. Using HARPS spectra, \cite{2010A&A...515A.106F} demonstrated that the visible atmospheric lines recorded by HARPS are RV references that are stable down to better than 10\,m/s over timescales of
ten years. Moreover, the authors showed that the shape of the O$_2$ lines could be parametrized using simple line profile and wind models, a result confirmed using radiosonde data \citep{2012MNRAS.420.2874F}. Several studies have used atmospheric lines as reference and modeled it for a precise removal, showing this to be an interesting and promising choice (see introduction and references therein). However, the imperfect removal of the imprint of atmospheric lines will introduce RV systematics on its own, and we demonstrated in Sect.\,\ref{AtmFlag} that this is a complex problem that very likely has a strong impact on the final achievable precision.

\subsection{Stellar models}

We here chose to use synthetic spectra to perform the simulations. While empirical spectral libraries are often preferred because atomic and molecular lines have the correct strengths, shapes, and positions, synthetic spectra present themselves as an attractive alternative, essentially because they can provide a wider range of wavelength ranges, resolutions, and atmospheric stellar parameters. This remains true even if they are naturally limited by the incompleteness of their line lists, inaccuracy of line parameters, and simplifying assumptions necessary for numerical computations. The ATLAS9 spectra based on Kurucz atmospheres \citep{1979ApJS...40....1K}, have been widely used for solar-type stars, but they do not cover the entire M dwarf range (only down to 3000\,K with steps of 250\,K). The PHOENIX model atmosphere code \citep{1999JCoAM.109...41H} has been used since it was first developed to create synthetic spectra for the low-mass regime \citep[e.g.,][]{1995ApJ...445..433A}. The most recent synthetic libraries based on PHOENIX model atmospheres are the BT-Settl models \citep{2012RSPTA.370.2765A} and the PHOENIX-ACES models \citep[Goettingen spectral library,][]{2013A&A...553A...6H}. There are two main differences between these two models The first is the version of the PHOENIX model atmospheres used: BT-Settl models use the PHOENIX v.15.5 atmospheres, while the PHOENIX-ACES models uses PHOENIX v.16, which employs a different equation of state (the astrophysical chemical equilibrium solver EOS). The second, and arguably most important, difference is that BT-Settl models do account for dust settling, and therefore they extend toward the very low mass and brown dwarf regime, while the PHOENIX-ACES models do not extend to temperatures lower than 2300\,K. In terms of synthetic spectrum characteristics, the PHOENIX-ACES models have a constant grid step in each wavelength range covered, while BT-Settl models do not have a constant wavelength sampling (extra wavelength points are created to better account for opacity). Both synthetic spectral libraries use spherical symmetry approximations and current solar element abundances \citep{2009ARA&A..47..481A}; the newest version of BT-SETTL also includes abundances from \cite{2011SoPh..268..255C}. 

As stated before, the difference in the treatment of dust is the main difference between the two models. For objects with temperatures slightly higher than 2800 K, it is suggested that the impact of the presence of dust clouds should be considered. This means that the results obtained with the spectra of 2800K and 2600K might be affected by the non-inclusion of dust in the models used. However, for the cooler stars, the influence of dust clouds in their spectra may not be evident because the efficiency of dust formation is weak at these temperatures (2800K-2600K) and therefore the opacity generated by these grains is quite low \citep{2011A&A...529A..44W}. The presence of dust clouds will mostly affect the optical spectrum of the cooler M dwarfs by removing the TiO and VO bands, as suggested by the DRIFT-PHOENIX models, which include dust clouds, by the Drift code \citep{2008A&A...485..547H}, in the calculation of the thermal structures and radiation field of the PHOENIX models \citep{2009A&A...506.1367W}.

It would be extremely interesting to compare the spectra obtained from BT-SETTL and PHOENIX-ACES. A significant difference between the output spectra can lead to different appreciable RV precision. However, the BT-SETTL spectra are produced at a lower resolution, which would place them on uneven footing for a comparison with PHOENIX-ACES. 

\subsection{Comparison with previous works}

It is very interesting to compare our results with those previously published. \cite{2010ApJ...710..432R} recreated spectra of M3, M6, and M9 in the JHK bands as observed at the same resolutions as considered here. There is no information on any broadening of the star due to its own rotation, therefore we assume this broadening mechanism was not considered or applied. The fact that the authors normalized their spectra relative to the Y band (a more central wavelength to their study) and used a slightly different way to evaluate spectral information leads us to compare only the relative gain between bands and not in absolute values. 
On M3 dwarfs we obtain the same relative precision: Y, H and J, in order of decreasing precision. Our conclusions also agree for M6 and M9 spectral type.

A close look at the M3 results shows, however, that while the relative order of the bands is the same, the ratio between the precision achieved on the different bands is slightly different. The low precision reported by \cite{2010ApJ...710..432R} in the J band is in part due to discarding a significant wavelength domain affected by atmospheric features. A lower precision in the H band (in ratio, when compared to ours) is probably due to the discarding of a larger portion of the spectra than we do when applying our condition 2. \cite{2010ApJ...710..432R} reported a ``telluric loss'' on the wavelength for the bands Y, J, and H of 19\%, 55\%, and 46\%, respectively. Considering an evenly spaced grid in wavelength, we calculate a loss of 8\%, 46\%, and 20\%; these are lower values than theirs; the value on the H band is not only relatively high but significantly different, and since \cite{2010ApJ...710..432R} discarded a larger domain of the spectra, their lower precision can be explained by this. However, the most striking point is in the expected loss for the K band (the authors did not analyze this band): \cite{2010ApJ...710..432R} reported a value of 80\%, while we obtain a value of 45\%. As we showed before, however, such differences in total discarded range can arise depending not only on the fidelity of the atmospheric model, but also on the sampling and identification of the shallow ubiquitous 2\% lines seen at R\,=\,100 000. 

A comparison with \cite{2011A&A...532A..31R} is also in order. The authors reproduced the spectrum of an M9.5 dwarf with a rotational velocity of 5\,km/s as observed by spectrographs with resolutions of 60\,000 and 80\,000 as seen in the YJHK bands. The case of perfect line removal was considered, but without amplifying the scatter due to low telluric absorption (corresponding to our condition 1) and after discarding polluted areas (condition 2). For condition 1, our results are very similar for R\,=\,80\,000; but for R\,=\,60\,000, \cite{2011A&A...532A..31R} reported that the highest precision is given by the K band, and puzzlingly even higher than its higher resolution counterpart. For condition 2 we reach similar relative precision values, with the difference that \cite{2011A&A...532A..31R} reported a mildly higher relative loss in the K band than we found. This might be due to their atmospheric model having a different resolution (R\,=\,80 000), but it is hard to conclude with the information at hand.

When performing our analysis we considered using the visual counterpart of PHOENIX-ACES spectra to compare the performance deliverable using visible wavelengths with those of the infrared. Other than having to deal with minor aspects associated with how diffuse atmospheric absorption would be considered for the transmission criteria \citep[see, e.g., the Chapuis ozone bands in Fig.\,1 of ][]{2015A&A...576A..77S}, the comparison would be, by construction, unfair. This is so because the work of line identification and modeling in the visible has reached a level of completeness and perfection that is not yet possible in the near-infrared due to the lack of high-resolution spectrographs. The lack of lines and mismatch between observed spectra and those used here might even be the strongest limitation to our work. Artigau et al. ({\it in prep}) show that the quality of the modeling clearly depends on wavelength: it becomes worse for the redder bands and is associated with a larger deficit of lines. This effect will lead to an increase of the information content as a function of wavelength, which will essentially favor the H and K bands. Unfortunately, the magnitude of the effect and RV impact is very hard to estimate and will essentially remain unknown until the advent of high-resolution spectrographs.

\subsection{Properties of the M-dwarf population in the solar neighborhood}

To set these results into context, we need to see them in light of the properties of the M dwarfs around us.
It is well known that M dwarfs constitute the vast majority of stars in the solar neighborhood, corresponding to more than 70\% of the total population \citep{2006AJ....132.2360H}, a number that is subject to revision as more of these stars are detected \citep[e.g.,][]{2015AJ....149....5W}. Indeed, within 5 parsecs, M-dwarf stars outnumber by roughly five times the total number of FGK stars in the most complete volume-limited sample of stellar objects \citep{1969PASP...81....5V, 2013AJ....146...99C}. The total number of fainter degenerate objects remains unknown, as the recent discoveries of nearby brown dwarfs within $\sim$2 pc demonstrate \citep{2013ApJ...767L...1L, 2014ApJ...786L..18L}. Two thirds of the 50 M dwarfs within 5 parsecs are mid-M dwarfs (M4, M5, and M6) with visual magnitudes in the range 9.5\,$<$\,V\,$<$\,17.5.  
 
\begin{figure}

\includegraphics[width=6.5cm, angle=270]{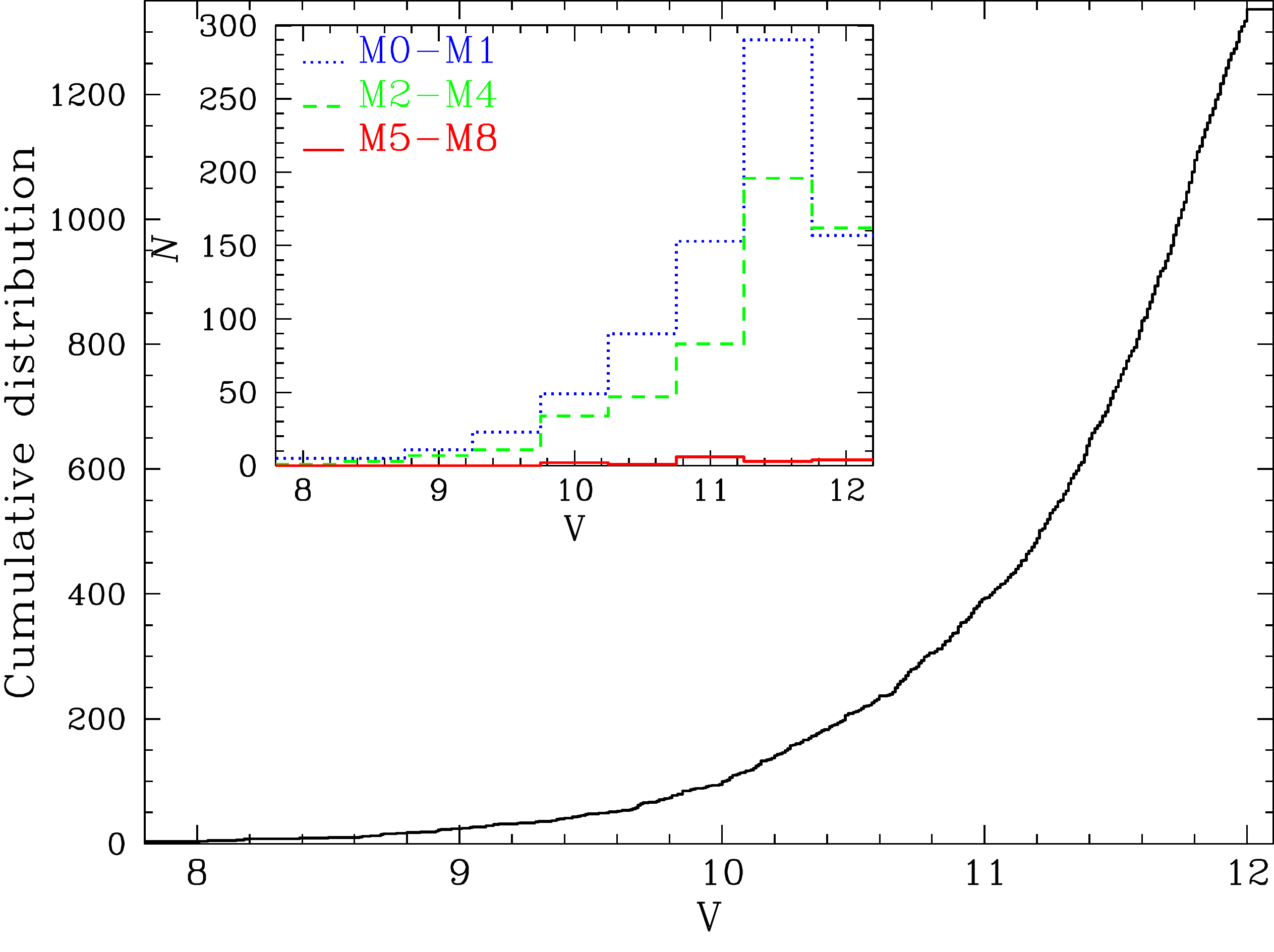}

\caption{Cumulative distribution of V magnitude for M-dwarf stars in the CONCH-SHELL catalog, as described in the text. {\it Inset}: distribution of V magnitude as a function of spectral type classes.}\label{Mdwarf_dist}

\end{figure}

Most RV planetary searches target relatively bright stars (V $<$ 12) in the solar vicinity, and as such the surveyed population is not representative of the whole M-dwarf population, since a cut in apparent magnitude will exclude stars in a way that depends on the spectral types. The final distribution will then differ dramatically from the population of stars in the solar neighborhood as outlined by the stellar mass function \citep[e.g., of][]{2010AJ....139.2679B}. To obtain a general picture of the M-dwarf sample targeted by RV searches we selected potentially observable stars from the CONCH-SHELL catalog of \cite{2014MNRAS.443.2561G}. This all-sky catalog of nearby (d $<$ 50 pc), bright (J $<$ 9) late-type dwarf stars reaches an estimated completeness of 98.6\% in the northern sky and 88.4\% in the southern hemisphere, making it the most complete catalog of its type. It lists a total of 2970 stars. Applying the conditions $T_{\mathrm{eff}}$\,$\leq$\,4000\,K, V\,$\leq$\,12 and spectral type classification of an M dwarf, we reached a total of 1345 stars. The cumulative distribution of the V magnitude is presented in Fig.\,\ref{Mdwarf_dist}, where we plot the magnitude distribution for different spectral subtypes
in the inset. As noted above, the sample is strongly biased against very late-type stars (of which only 13 stars were listed) due to the magnitude cutoff of the catalog.

In this work we focused on the achievable RV precision on a given M-dwarf as dictated exclusively by their spectroscopic properties, in what is a very simplified picture. However, stellar phenomena can create parasite RV signals that can hamper planetary detection capabilities. At the m/s level studied in this work, three mechanisms can contribute to the RV signal budget: oscillations, granulation, and stellar activity. For the M-dwarf regime, stellar oscillations are probably the least important of the three; their amplitude is proportional to the ratio of stellar luminosity over stellar mass \citep[e.g.,][]{2004SoPh..220..137C}, and while predicted theoretically, it has never been successfully detected \citep{2012MNRAS.419L..44R, 2015MNRAS.446.2613R}. The amplitude of granulation signals is proportional to the velocity of convection and is also much lower in M dwarfs than in their solar-type counterparts, which leaves stellar activity as the third mechanism. It has been repeatedly
shown that stellar activity can be a nuisance both by introducing signals that can be mistaken for that of a planet and by contributing to the noise as a stochastic process \citep[e.g.,][]{2001A&A...379..279Q, 2008A&A...489L...9H, 2010A&A...513L...8F}. As a star rotates, active regions induce an RV variation by breaking the symmetry of the stellar disk in the flux distribution and convective blueshift inhibition \citep[e.g.,][]{1997ApJ...485..319S, 2011A&A...527A..82D}. To characterize the impact of stellar activity in RV measurements, we need then to know not only its spot or plages coverage, but the rotational velocity of the host.  

Several studies have shown that both the average rotational velocity and average activity level increase toward later-type M dwarfs. The data of \cite{2012AJ....143...93R} indicate that the fraction of fast rotators ($v.\sin{i}$\,$\geq$\,3 km/s) increases from about 5\% for M0-M2 types to about 45\% for M4 type stars. The same data lead to an average $v.\sin{i}$ value of $\approx$3.5 km/s for M0-M1 and $\approx$4.5 km/s for M2-M4 type stars. It is important to note, however, that for a very large portion of the stars used to derive these values, the rotational velocities were upper limits (with the detection thresholds for the measurements ranging from 3 to 4 km/s, depending on the spectrographs used). For later-type M dwarfs we calculated average values of $v.\sin{i}$ by using the sample and literature compilation of \cite{2009ApJ...704..975J}. This data give similar average values of $v.\sin{i}$ for M0 (3.2 km/s) and M3 (5.7\,km/s) and point to much higher values for later-type objects: 9.2\,km/s for M5-M7 type stars and about 19\,km/s for M8-M9. However, these numbers should be used with caution because the number of stars in the last spectral bin is small (only 18 stars) and comes from an inhomogeneous literature compilation containing data coming from young, active, and fast-rotating stars. But it is important to note that several studies point toward a bi-modal $v.\sin{i}$ distribution on M dwarfs. The recent results on Kepler data \citep{2013MNRAS.432.1203M, 2014ApJS..211...24M} show that the distribution of rotation periods of M dwarfs is bimodal, being characterized by two groups with periods in the range 10-25 and 25-80\,days, with peaks at $\sim$19 and $\sim$33\,days. The same works also found evidence of an increase of rotational period with decreasing mass, contrary to what we see for K dwarfs. The two groups are associated with different formation epochs and regions, but further evidence is required for this to be confirmed. This modality in rotational period will lead to a bimodality in $v.\sin{i}$, but the average values and their spread will depend on the details of how the rotational periods depend on spectral type. For the periods of 19 and 33\,days and assuming a stellar radius if 0.4$R_\odot$ (M2), we obtain a rotational velocity of 1\,km/s and 0.6\,km/s, respectively, which is below the resolution of current spectrographs. Interestingly,  \cite{2012AJ....143...93R} already indicated bimodality in $v.\sin{i}$ in M4-5, as can be seen in their Fig.\,7. 

In a similar way, the stellar chromospheric activity is higher for later-type M stars because of their substantially longer activity lifetimes. \cite{2012AJ....143...93R} showed that the fraction of active M stars (i.e., with $H_{\mathrm{\alpha}}$ detected in emission) increases from 5\% for M0 to 55\% for M4 dwarfs. Similar values for early and mid-type M dwarfs were obtained by \cite{2013AJ....145..102L} and for the significantly more extended LAMOST survey analyzed by \cite{2014AJ....147...33Y}. This is less clear for later-type stars. \cite{2013AJ....145..102L} calculated that the fraction of active stars increases to about 90\% for M6 type stars, with a somewhat lower fraction of active stars reported by \cite{2011AJ....141...97W} from the analysis of 71 000 M dwarfs from the SDSS DR7. The lower values obtained ($\approx$5\% for M0, 20\% for M4, $\approx$60\% for M6, and $\approx$80\% for M9-type dwarfs) probably  arise because the distance of the stars in the \cite{2011AJ....141...97W} sample to the Galactic plane is on average larger than those of \cite{2013AJ....145..102L} and \cite{2012AJ....143...93R}, which biases the sample toward older stars.

\begin{figure*}

\centering

\includegraphics[width=14cm]{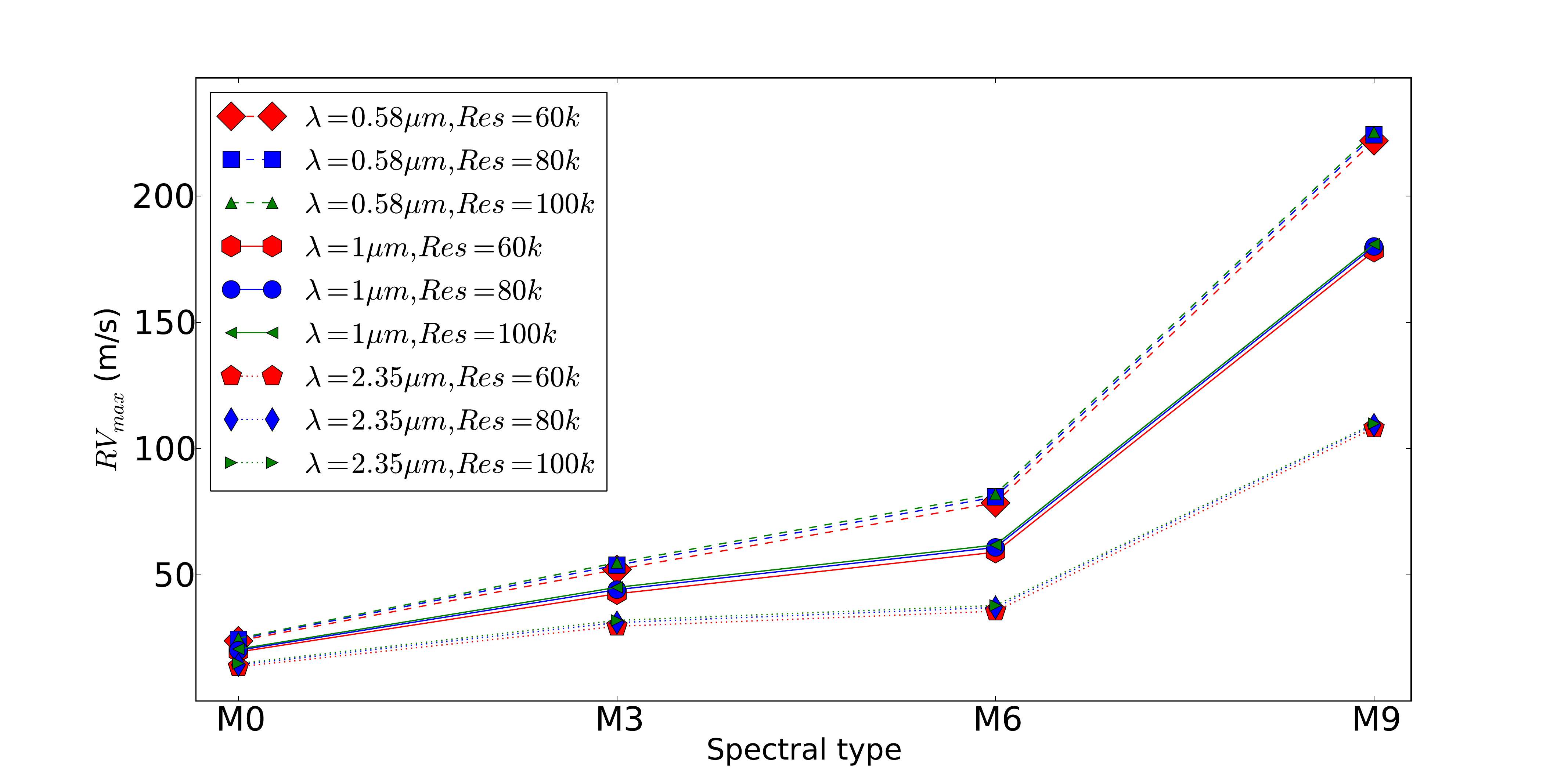}

\caption{Peak-to-peak amplitude of the RV impact introduced by a 1\% spot simulated by SOAP2.0 on M dwarfs of different spectral types and associated $v.\sin{i}$, of value 3.2, 5.7, 9.2, and 19 km/s for M0, M3, M6, and M9, respectively. The different symbols and line styles represent different observing wavelengths, and the different colors represent different observing resolutions.}\label{SOAP_Fig1}

\end{figure*}

These values can be translated into potential RV-precision impact. We used the software SOAP 2.0 \citep{2014ApJ...796..132D} to simulate the amplitude of an activity-induced RV signal on an M-dwarf population with the properties described above. Unfortunately, and while we know late-type stars experience increased activity, the value of this increase is still a matter of debate and cannot be translated directly into a spot filling factor. Therefore, we considered 1\%, the highest value during the solar activity cycle \citep{2003A&ARv..11..153S}, for all the spectral types. For the spot temperature difference relative to the photosphere we considered 500\,K for spectral types M0 and M3, and 250\,K for spectral types M6 and M9, following \cite{2015MNRAS.448.3053A} and \cite{2005LRSP....2....8B}. For the stellar properties of the different dwarfs we considered the temperatures listed in Sect.\,\ref{simuls} and radii of 0.62, 0.39, 0.15, and 0.08\,$R_{\odot}$ for M0, M3, M6, and M9 dwarfs, respectively \citep{2009ApJ...698..519K}. In line with the procedure described in Sect.\,\ref{simuls}, we considered a linear limb-darkening coefficient $\epsilon$ of 0.6\footnote{We note, however, that some studies considered wavelength-dependent limb-darkening \citep[e.g.,][]{2014A&A...568A..99O}, but for the level of precision we require for this discussion we consider this to be unnecessary.}. The $v.\sin{i}$ assigned to each spectral type was taken from the values presented in the previous paragraph, being of 3.2, 5.7, 9.2, and 19\,km/s for M0, M3, M6, and M9, respectively. The impact on the RV is presented in Fig.\,\ref{SOAP_Fig1}.

We can see that there is a significant increase in RV signal amplitude toward later-type stars. It is also worth mentioning that in the K band the RV measurements are much less affected by the activity features than in the Y band and visible. The impact of the presence of a spot on RV is expected to depend on the contrast between the stellar spot and the surrounding photospheric temperature \citep[e.g.,][]{2006ApJ...644L..75M, 2010A&A...511A..55F}. This behavior is also found in such a simple model as ours, even though we caution that Zeeman broadening on spectral lines may occur, of which we do not have precise quantitative knowledge, but which is expected to show an opposite wavelength dependence \citep{2013A&A...552A.103R, 2014MNRAS.443.2599H}. Observing at different high-resolution values does not seem to have a significant impact on the measured RV value. Despite of all the uncertainties, our simple analysis shows that there is an appreciable difference when we move from M0 to M9 stars, on which we measure a much higher activity RV impact. For a comprehensive analysis of this aspect, we refer to \cite{2011MNRAS.412.1599B} and \cite{2015MNRAS.448.3053A}, who made a set of hypotheses on spot properties and derived detection limits from them. 
As stated before, there are virtually no comprehensive statistics about how spot coverage, and in particular the spot filling factor, vary in the M-dwarf spectral range, and we refer to \cite{2015MNRAS.448.3053A} for a comprehensive review of the literature on these two aspects. To explore the possibility of larger spot coverage, we varied the filling factor that characterizes the active region. If the filling factor is increased by a factor of ten, to 10\%, the RV amplitude is increased by a factor of ten as well, as shown in Fig.\,6 of \cite{2012A&A...545A.109B}. Since activity is expected to be related to spot coverage, the increasing activity level from M0-M3 to M6-M9 is expected to clarify the divide and different impact on RV between early and late-type stars.

A final aspect to consider is how the different absolute magnitudes of M dwarfs with different spectral types affect the measurable RV. To illustrate this effect we considered the $K_s$ magnitudes from Table\,2 of \cite{2014A&A...571A..36R}: for an M dwarf at a distance of 60\,pc we have for spectral types of M0, M3, M6, and M9 absolute magnitudes of 9.1, 10.6, 12.9, and 14.5. From the previously mentioned formula of \cite{1992ESOC...40..275H} we have that the RV precision is $\propto\sqrt{F_{meas}}\,\propto\sqrt{10^{-m/2.5}}$, so that by normalizing the RV precision on an M0 to 1\,m/s, we have a precision of 1.0, 2.0, 5.7, and 12\,m/s for M0, M3, M6 and M9, respectively. This means that if M dwarfs are observed at the same distance, the relative gain in precision discussed in this paper can be much smaller than the effect of relative flux. However, it is unrealistic to assume that the different distribution of observed M dwarfs places them at the same average distance from us, and thus on an even footing for observation. Therefore we prefer to present the detailed values for normalized spectra, but add a word of caution. 
For a study that parametrizes the variable factors, such as telescope size, transmission, and other parameters, keeping the exposure time fixed to evaluate the best hosts for exoplanetary detection, we refer to  \cite{2013PASP..125..240B}.

In summary, our knowledge on the M-dwarfs in the solar neighborhood is much more complete for early-M dwarfs, around which exoplanets are also easier to detect by RV due to lower $v.\sin{i}$ values and reduced activity when compared with their later counterparts. A precise assessment of the impact of activity throughout the M-dwarf spectral class requires detailed knowledge of spot coverage properties, which is inaccessible to us. Because spot coverage is expected to increase with activity, however, and activity has been shown to be higher for later-type M dwarfs, the effect on these hosts is very probably stronger than on early-M stars and will accentuate the difference as presented in Fig.\,\ref{SOAP_Fig1}. In spite of these accrued difficulties, later M-dwarfs are the most abundant stars and the ones for which transit and RV signals have a higher amplitude. Relatively little is known about planet formation processes around these stars, but they remain very promising candidates for surveys. While the number of bright (V\,$<$\,12) early-M dwarfs fully justifies the existence of a dedicated spectrograph, such an instrument should therefore also be able to measure precise RV on the lowest-mass and faintest stars in our galaxy.   

\subsection{Achieving the highest RV precision and designing dedicated spectrographs}

The work presented here can be of value for the design and development of nIR spectrographs. The preceding sections showed that the bands Z,Y, and H allow the most precise RV measurements for M0-M3 dwarfs (with relative merits depending on the specific conditions), while the difference in precision between bands becomes very small for M6-M9 stars. The difference between bands when correcting for the atmosphere transmission, however, is never larger than a factor of three. For later M stars, the precision delivered by a spectra with the same S/N increases because there are more features.

Interestingly, the dichotomy present at the level of spectral features and reported here also exists at the level of other stellar characteristics, such as average activity level, rotation, spot number, and magnitude.
An important result can be achieved by merging the information from these two aspects. A spectrograph aiming at observing early-type M dwarfs will take great advantage by observing in the ZY bands and extending it to H band if possible. This point was squarely put into practice in the design of HZPF \citep{2012SPIE.8446E..1SM}, a spectrograph that with a relatively low-cost design will be able to reach high precision.
On the other hand, a spectrograph that aims at studying late-M dwarfs in the M6-M9 domain will benefit from extending its wavelength domain as much as possible and should cover the reddest wavelengths. The extended wavelength will allow reaching a higher precision on these fainter hosts within a reasonable integration time; the smooth dependence on wavelength reported for non-rotating late-type stars shows that an increased precision can be obtained by observing up to the red K band. Moreover, the longer wavelength arm, aided by activity-correction capabilities (such as polarimetry) will contribute strongly to reaching a competitive precision on the active late-type stars. These were exactly the technical choices made for the SPIRou spectrograph \citep{2014SPIE.9147E..15A}.

\begin{figure*}
\centering
\includegraphics[width=14cm]{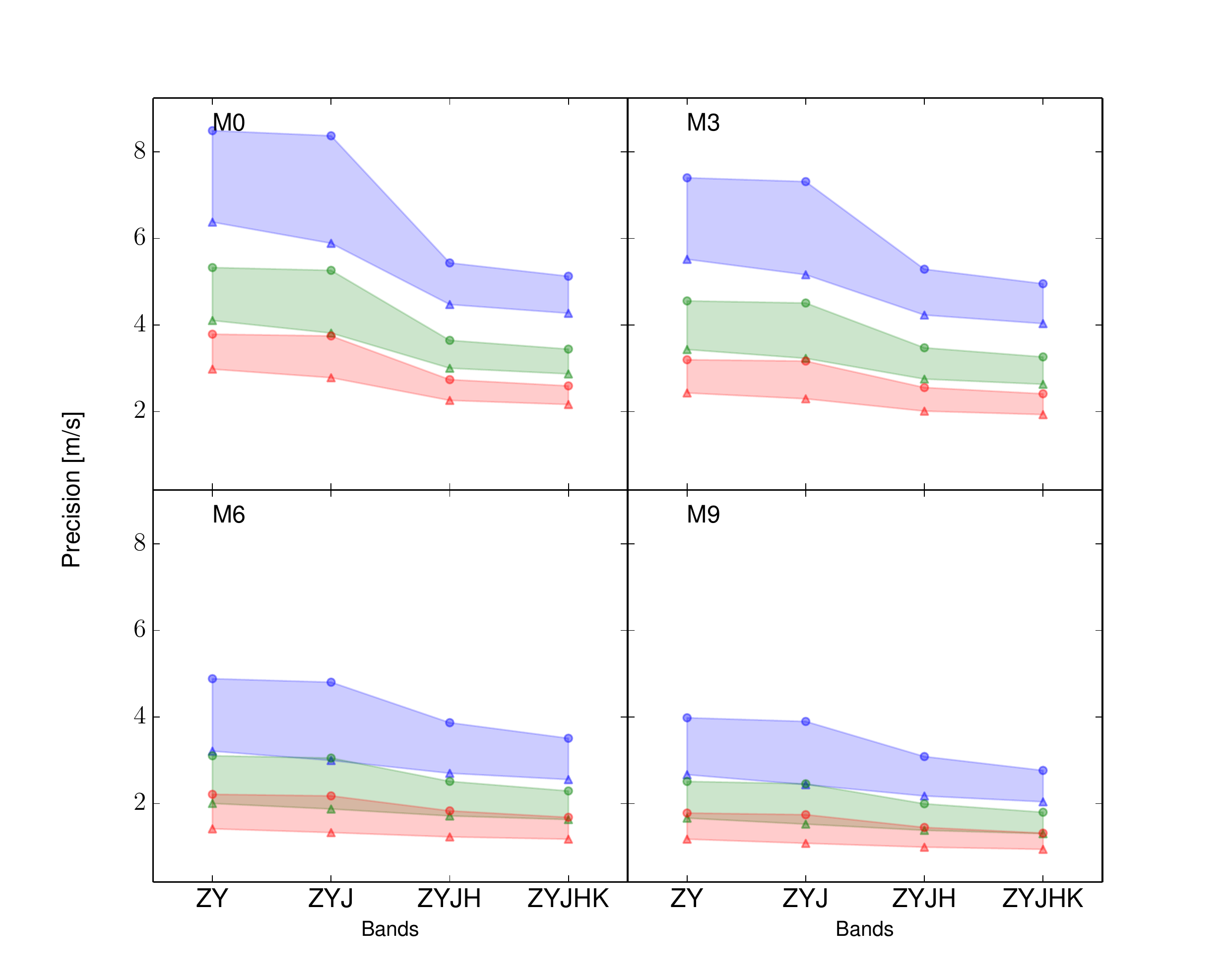}

\caption{Precision achieved when considering simultaneous observation on different spectral bands for stars with a  rotational velocity of $v.\sin{i}$\,=\,1.0\,km/s and spectral types M0, M3, M6, and M9. The filled area represents the values within the limits set by conditions 2 ({\it circles}) and 3 ({\it triangles}); blue, green and red represent the results obtained for resolutions of 60\,000, 80\,000, and 100\,000, respectively. The spectra were normalized to have a S/N of 100 per resolution element as measured at the center of the J band (see Sect.\,\ref{simuls} for details).}\label{TotalRV}

\end{figure*}

To better grasp the results delivered by observing with several bands simultaneously, we consider the error from performing the weighted average from the different bands. In Fig.\,\ref{TotalRV} we plot the results of this calculation when considering progressively redder bands. We can picture the high precision delivered by the ZY bands alone and the gain achieved by including the H band in the high-resolution observations of M0-M3 stars. We also note that M6 and M9 deliver an higher precision overall than their hotter counterparts. It is also clear that for these hosts observing in the different bands leads to similar precisions, and the coverage of the full near-infrared spectrum allows reaching the 1 m/s level on a spectrum with S/N\,=\,100. It is also interesting to note how correcting or not correcting for the atmosphere leads to different results.

We here assumed the throughput to be the same for the different scenarios, in particular for the different spectrograph resolutions. It is important to note that there is no universal relation between optical efficiency and spectral resolution. In principle, it is possible to design an arbitrarily efficient instrument, even at very high spectral resolution, as long as the financial and technical budgets are unlimited. Furthermore, we recall that both optical efficiency and spectral resolution contribute to the photon-noise budget. As long as the stellar lines are not (fully) resolved, the photon-noise-limited RV precision scales with R$^{1.5}$ \citep[e.g.,][]{1992ESOC...40..275H}, and therefore it might be advantageous to lose photons, at fixed instrument costs, for the sake of higher spectral resolution and thus RV precision. It is the objective of this work to quantify the impact of spectral resolution on the RV measurement precision and to help decide in the trade-off analysis between slit-efficiency and spectral resolution for future IR spectrographs.

\section{Conclusions}\label{sec:Conclusions}

We evaluated the RV content of ZYJHK bands for M-dwarfs of spectral types M0, M3, M6, and M9. We considered three values for $v.\sin{i}$, of 1.0, 5.0, and 10.0\,km/s and resolutions of 60\,000, 80\,000, and 100\,000. Our study was the first to analyze the content of the Y band and cover a wide range of spectral types and rotational velocities. To evaluate the information content of these spectra, we followed the procedure of \cite{2001A&A...374..733B}, considering for the analysis both only areas not polluted by deep absorption features and the whole spectrum, but after boosting the noise as a consequence of an atmospheric transmission lower than one. The former condition has been explored before, but the latter was proposed by us for the first time and emulates the result obtained if a spectrum is perfectly corrected for the transmission of the Earth's atmosphere. The main conclusions of our work are
listed below.

\begin{enumerate}
 \item The spectral type is the most important parameter influencing the precision achievable on the different bands. For M0-M3 stars the Z, Y, and H bands are the most favorable, with the relative merit depending on the other parameters considered. However, for M6-M9 stars the different bands provide the same precision within a factor of $\sim$1.5 and show a reduced dependence on wavelength when compared with younger counterparts. 
 
 \item An M3-M6 star will deliver a precision $\text{about}$ twice better than an M0-M3 star observed with the same S/N, which is easily explained by the increase in the number of stellar features as the effective temperature decreases.
 
 \item The details of modeling the Earth's atmosphere and interpreting the results have a significant impact on the wavelength regions that are discarded when setting a limit threshold at 2-3\%. The resolution element sampling on the observed spectra plays an important role on the atmospheric transmission characterization. As a result of the multiparameter nature of the problem, it is very difficult to precisely quantify the impact of the telluric lines on the RV precision, but it is an important limiting factor to the achievable RV precision.

 \item A fraction of the difference in achievable RV when observing with the different bands comes from the contamination by tellurics and subsequent rejection of the polluted spectral regions. The fine-tuning of the rejection criteria can lead to different final precision and might overestimate the real difference between bands even after a poor correction. The lower values on the difference in precision between the different bands reported by us are due in part to the identification and isolation of this factor.
  
\end{enumerate}

Our work can be valuable for the design and development of nIR spectrographs and allow tailoring them to specific scientific goals. As an example, we concluded that the HZPF spectrograph has an ideal design for studying bright M0-M3 stars (and at a relatively low cost), while SPIRou is particularly adapted for studying M6-M9 dwarfs and 
extracting the most precise RVs from these active and faint stars.

Our study made use of the most recent PHOENIX-ACES models to create the observed spectrum. Of the different hypotheses and limitations that affect the results derived from this study, the choice of the models and its limitations are probably the most important. A cross-comparison between models and the compilation of a high-resolution and high-fidelity near-infrared library for M dwarfs would allow understanding the uncertainties on atmospheric models and the completeness of atomic and molecular databases and finding the best solution for the RV information content. 


\begin{acknowledgements}
This work was supported by Funda\c{c}\~ao para a Ci\^encia e a Tecnologia (FCT) through the research grant UID/FIS/04434/2013. PF and NCS acknowledge support by Funda\c{c}\~ao para a Ci\^encia e a Tecnologia (FCT) through Investigador FCT contracts of reference IF/01037/2013 and IF/00169/2012, respectively, and POPH/FSE (EC) by FEDER funding through the program ``Programa Operacional de Factores de Competitividade - COMPETE''. PF further acknowledges support from Funda\c{c}\~ao para a Ci\^encia e a Tecnologia (FCT) in the form of an exploratory project of reference IF/01037/2013CP1191/CT0001. VA and BR-A acknowledge the support from the Funda\c{c}\~ao para a Ci\^encia e a Tecnologia, FCT (Portugal) in the form of the fellowships SFRH/BPD/70574/2010 and SFRH/BPD/87776/2012, respectively. MO acknowledges support from Centro de Astrof\'{i}sica da Universidade do Porto through the grant of reference of CAUP-15/2014-BDP. JJN acknowledges support from FCT in the form of a ``PhD::Space'' (PD/00040/2012) network doctoral grant, of reference PD/BD/52700/2014. PF thanks Jean-Loup Bertaux and the team behind the TAPAS platform for developing such a user friendly interface and their availability to discuss and improve upon it. PF also thanks Jo\~{a}o Faria and Paulo Peixoto for a limitless supply of python and informatic wizardry. Finally, PF thanks Jean-Fran\c{c}ois Donati and \'Etienne Artigau for their comments on the manuscript and helpful discussions. We thank the anonymous referee for a careful reading and valuable comments on the paper.  

\end{acknowledgements}

\bibliographystyle{aa} 
\bibliography{Mybibliog,extra} 

\begin{appendix}

\section{Detailed results of the simulations}\label{AppendixRes}

The results of the simulations as described in the paper are presented here. We calculated the precision considering three
conditions: 
\begin{enumerate}
 \item observations are available over the whole spectral band;
 \item pixels at less than 30\,km/s from telluric absorption features deeper than 2\% are discarded;
 \item considering that the variance represented in the denominator, i.e. the photon noise contribution to the spectrum, is amplified by the telluric correction, and is given by the measured photon count $A_{0}(i)$ divided by the telluric transmission at that wavelength $T(\lambda(i))^2$.
\end{enumerate}

\begin{longtable}{cccc}

\caption{Results of the simulations described in Sect.\,\ref{simuls}.}\label{TableData}
\\
\hline\hline
Simulation & $\sigma_{RV}$(Cond.\,1) & $\sigma_{RV}$(Cond.\,2) & $\sigma_{RV}$(Cond.\,3) \\ 
(SpTp - Band - $v.sin{i}$ - R) & [m/s]& [m/s] & [m/s]\\
 \hline
 \endfirsthead
\caption{continued.}\\
\hline\hline
Simulation & $\sigma_{RV}$(Cond.\,1) & $\sigma_{RV}$(Cond.\,2) & $\sigma_{RV}$(Cond.\,3) \\ 
\hline
\endhead
\hline
\endfoot

M0-Z-1.0-60k & 8.9       & 26.1  & 9.3 \\
M0-Z-1.0-80k & 6.0       & 17.1  & 6.2 \\
M0-Z-1.0-100k & 4.5      & 12.8  & 4.6 \\
M0-Z-5.0-60k & 13.6      & 38.9  & 14.0 \\
M0-Z-5.0-80k & 10.6      & 30.5  & 10.9 \\
M0-Z-5.0-100k & 9.0      & 25.7  & 9.3 \\
M0-Z-10.0-60k & 23.9     & 60.8  & 24.6 \\
M0-Z-10.0-80k & 19.9     & 50.6  & 20.5 \\
M0-Z-10.0-100k & 17.4    & 44.2  & 18.0 \\
M0-Y-1.0-60k & 8.7       & 9.0   & 8.8 \\
M0-Y-1.0-80k & 5.4       & 5.6   & 5.5 \\
M0-Y-1.0-100k & 3.8      & 4.0   & 3.9 \\
M0-Y-5.0-60k & 14.6      & 15.1  & 14.8 \\
M0-Y-5.0-80k & 11.1      & 11.4  & 11.2 \\
M0-Y-5.0-100k & 9.1      & 9.4   & 9.2 \\
M0-Y-10.0-60k & 29.6     & 30.7  & 30.0 \\
M0-Y-10.0-80k & 24.2     & 25.0  & 24.4 \\
M0-Y-10.0-100k & 20.7    & 21.4  & 20.9 \\
M0-J-1.0-60k & 14.5      & 50.3  & 15.3 \\
M0-J-1.0-80k & 9.8       & 33.9  & 10.3 \\
M0-J-1.0-100k & 7.4      & 25.6  & 7.8 \\
M0-J-5.0-60k & 21.6      & 72.3  & 22.9 \\
M0-J-5.0-80k & 16.9      & 56.1  & 17.9 \\
M0-J-5.0-100k & 14.3     & 47.3  & 15.1 \\
M0-J-10.0-60k & 37.4     & 115.1         & 39.7 \\
M0-J-10.0-80k & 31.0     & 95.2  & 33.0 \\
M0-J-10.0-100k & 27.2    & 83.5  & 28.9 \\
M0-H-1.0-60k & 6.7       & 7.1   & 6.9 \\
M0-H-1.0-80k & 4.8       & 5.1   & 4.9 \\
M0-H-1.0-100k & 3.8      & 4.0   & 3.9 \\
M0-H-5.0-60k & 9.7       & 10.2  & 9.9 \\
M0-H-5.0-80k & 7.6       & 8.0   & 7.7 \\
M0-H-5.0-100k & 6.4      & 6.8   & 6.6 \\
M0-H-10.0-60k & 17.0     & 17.9  & 17.4 \\
M0-H-10.0-80k & 14.1     & 14.9  & 14.4 \\
M0-H-10.0-100k & 12.3    & 13.0  & 12.6 \\
M0-K-1.0-60k & 13.4      & 15.5  & 14.4 \\
M0-K-1.0-80k & 9.2       & 10.5  & 9.8 \\
M0-K-1.0-100k & 7.1      & 8.1   & 7.6 \\
M0-K-5.0-60k & 20.7      & 23.9  & 22.1 \\
M0-K-5.0-80k & 16.0      & 18.4  & 17.1 \\
M0-K-5.0-100k & 13.5     & 15.5  & 14.4 \\
M0-K-10.0-60k & 38.3     & 44.2  & 40.9 \\
M0-K-10.0-80k & 31.7     & 36.6  & 33.9 \\
M0-K-10.0-100k & 27.7    & 31.9  & 29.7 \\
M3-Z-1.0-60k & 7.7       & 22.6  & 7.9 \\
M3-Z-1.0-80k & 4.8       & 14.0  & 5.0 \\
M3-Z-1.0-100k & 3.5      & 10.0  & 3.6 \\
M3-Z-5.0-60k & 12.5      & 36.8  & 12.9 \\
M3-Z-5.0-80k & 9.6       & 28.3  & 9.9 \\
M3-Z-5.0-100k & 8.0      & 23.6  & 8.3 \\
M3-Z-10.0-60k & 23.8     & 61.4  & 24.6 \\
M3-Z-10.0-80k & 19.6     & 50.9  & 20.3 \\
M3-Z-10.0-100k & 17.1    & 44.3  & 17.7 \\
M3-Y-1.0-60k & 7.6       & 7.8   & 7.7 \\
M3-Y-1.0-80k & 4.7       & 4.8   & 4.7 \\
M3-Y-1.0-100k & 3.3      & 3.4   & 3.3 \\
M3-Y-5.0-60k & 13.0      & 13.4  & 13.2 \\
M3-Y-5.0-80k & 9.8       & 10.1  & 9.9 \\
M3-Y-5.0-100k & 8.1      & 8.3   & 8.2 \\
M3-Y-10.0-60k & 26.9     & 27.8  & 27.2 \\
M3-Y-10.0-80k & 21.9     & 22.6  & 22.1 \\
M3-Y-10.0-100k & 18.7    & 19.3  & 18.9 \\
M3-J-1.0-60k & 13.8      & 47.5  & 14.6 \\
M3-J-1.0-80k & 9.0       & 31.0  & 9.5 \\
M3-J-1.0-100k & 6.6      & 22.9  & 7.0 \\
M3-J-5.0-60k & 21.4      & 70.5  & 22.6 \\
M3-J-5.0-80k & 16.5      & 54.3  & 17.5 \\
M3-J-5.0-100k & 13.9     & 45.6  & 14.7 \\
M3-J-10.0-60k & 38.4     & 116.4         & 40.8 \\
M3-J-10.0-80k & 31.8     & 95.8  & 33.7 \\
M3-J-10.0-100k & 27.7    & 83.7  & 29.4 \\
M3-H-1.0-60k & 7.2       & 7.7   & 7.4 \\
M3-H-1.0-80k & 5.1       & 5.5   & 5.3 \\
M3-H-1.0-100k & 4.1      & 4.3   & 4.2 \\
M3-H-5.0-60k & 10.2      & 10.8  & 10.4 \\
M3-H-5.0-80k & 8.0       & 8.5   & 8.2 \\
M3-H-5.0-100k & 6.8      & 7.2   & 7.0 \\
M3-H-10.0-60k & 17.4     & 18.4  & 17.9 \\
M3-H-10.0-80k & 14.5     & 15.3  & 14.9 \\
M3-H-10.0-100k & 12.7    & 13.4  & 13.0 \\
M3-K-1.0-60k & 12.5      & 14.1  & 13.3 \\
M3-K-1.0-80k & 8.4       & 9.5   & 9.0 \\
M3-K-1.0-100k & 6.5      & 7.3   & 6.9 \\
M3-K-5.0-60k & 19.3      & 22.0  & 20.6 \\
M3-K-5.0-80k & 14.9      & 16.9  & 15.9 \\
M3-K-5.0-100k & 12.5     & 14.2  & 13.4 \\
M3-K-10.0-60k & 36.1     & 41.0  & 38.5 \\
M3-K-10.0-80k & 29.9     & 33.9  & 31.9 \\
M3-K-10.0-100k & 26.1    & 29.6  & 27.9 \\
M6-Z-1.0-60k & 3.9       & 11.3  & 4.0 \\
M6-Z-1.0-80k & 2.4       & 6.8   & 2.5 \\
M6-Z-1.0-100k & 1.7      & 4.7   & 1.7 \\
M6-Z-5.0-60k & 6.9       & 20.3  & 7.2 \\
M6-Z-5.0-80k & 5.2       & 15.3  & 5.4 \\
M6-Z-5.0-100k & 4.3      & 12.7  & 4.5 \\
M6-Z-10.0-60k & 14.4     & 39.5  & 15.0 \\
M6-Z-10.0-80k & 11.8     & 32.3  & 12.3 \\
M6-Z-10.0-100k & 10.2    & 28.0  & 10.6 \\
M6-Y-1.0-60k & 5.2       & 5.4   & 5.3 \\
M6-Y-1.0-80k & 3.4       & 3.5   & 3.4 \\
M6-Y-1.0-100k & 2.4      & 2.5   & 2.5 \\
M6-Y-5.0-60k & 8.3       & 8.6   & 8.4 \\
M6-Y-5.0-80k & 6.4       & 6.6   & 6.5 \\
M6-Y-5.0-100k & 5.4      & 5.5   & 5.5 \\
M6-Y-10.0-60k & 14.4     & 14.8  & 14.6 \\
M6-Y-10.0-80k & 11.9     & 12.3  & 12.1 \\
M6-Y-10.0-100k & 10.3    & 10.6  & 10.5 \\
M6-J-1.0-60k & 7.9       & 26.5  & 8.3 \\
M6-J-1.0-80k & 5.0       & 16.8  & 5.3 \\
M6-J-1.0-100k & 3.6      & 12.1  & 3.9 \\
M6-J-5.0-60k & 12.9      & 42.7  & 13.7 \\
M6-J-5.0-80k & 9.9       & 32.9  & 10.5 \\
M6-J-5.0-100k & 8.3      & 27.6  & 8.8 \\
M6-J-10.0-60k & 23.8     & 76.0  & 25.2 \\
M6-J-10.0-80k & 19.7     & 62.6  & 20.9 \\
M6-J-10.0-100k & 17.2    & 54.5  & 18.2 \\
M6-H-1.0-60k & 6.0       & 6.5   & 6.2 \\
M6-H-1.0-80k & 4.1       & 4.4   & 4.2 \\
M6-H-1.0-100k & 3.1      & 3.4   & 3.2 \\
M6-H-5.0-60k & 9.4       & 10.2  & 9.7 \\
M6-H-5.0-80k & 7.3       & 7.9   & 7.6 \\
M6-H-5.0-100k & 6.2      & 6.7   & 6.4 \\
M6-H-10.0-60k & 17.5     & 18.9  & 18.1 \\
M6-H-10.0-80k & 14.5     & 15.6  & 15.0 \\
M6-H-10.0-100k & 12.6    & 13.6  & 13.1 \\
M6-K-1.0-60k & 7.4       & 8.3   & 7.9 \\
M6-K-1.0-80k & 5.0       & 5.6   & 5.3 \\
M6-K-1.0-100k & 3.8      & 4.3   & 4.1 \\
M6-K-5.0-60k & 11.8      & 13.3  & 12.6 \\
M6-K-5.0-80k & 9.1       & 10.3  & 9.8 \\
M6-K-5.0-100k & 7.7      & 8.6   & 8.2 \\
M6-K-10.0-60k & 22.7     & 25.5  & 24.2 \\
M6-K-10.0-80k & 18.8     & 21.1  & 20.0 \\
M6-K-10.0-100k & 16.4    & 18.4  & 17.5 \\
M9-Z-1.0-60k & 3.3       & 9.4   & 3.4 \\
M9-Z-1.0-80k & 2.0       & 5.7   & 2.1 \\
M9-Z-1.0-100k & 1.4      & 4.0   & 1.5 \\
M9-Z-5.0-60k & 5.8       & 17.0  & 6.1 \\
M9-Z-5.0-80k & 4.4       & 12.8  & 4.6 \\
M9-Z-5.0-100k & 3.7      & 10.6  & 3.9 \\
M9-Z-10.0-60k & 12.2     & 33.6  & 12.7 \\
M9-Z-10.0-80k & 10.0     & 27.5  & 10.4 \\
M9-Z-10.0-100k & 8.6     & 23.8  & 9.0 \\
M9-Y-1.0-60k & 4.2       & 4.4   & 4.3 \\
M9-Y-1.0-80k & 2.7       & 2.8   & 2.7 \\
M9-Y-1.0-100k & 1.9      & 2.0   & 1.9 \\
M9-Y-5.0-60k & 6.9       & 7.1   & 7.0 \\
M9-Y-5.0-80k & 5.3       & 5.5   & 5.4 \\
M9-Y-5.0-100k & 4.5      & 4.6   & 4.5 \\
M9-Y-10.0-60k & 11.9     & 12.2  & 12.0 \\
M9-Y-10.0-80k & 9.9      & 10.1  & 10.0 \\
M9-Y-10.0-100k & 8.5     & 8.8   & 8.6 \\
M9-J-1.0-60k & 5.7       & 18.9  & 6.0 \\
M9-J-1.0-80k & 3.6       & 11.9  & 3.8 \\
M9-J-1.0-100k & 2.6      & 8.5   & 2.7 \\
M9-J-5.0-60k & 9.5       & 31.5  & 10.1 \\
M9-J-5.0-80k & 7.3       & 24.2  & 7.7 \\
M9-J-5.0-100k & 6.1      & 20.3  & 6.5 \\
M9-J-10.0-60k & 17.8     & 57.5  & 18.8 \\
M9-J-10.0-80k & 14.7     & 47.3  & 15.6 \\
M9-J-10.0-100k & 12.8    & 41.1  & 13.6 \\
M9-H-1.0-60k & 4.7       & 5.1   & 4.8 \\
M9-H-1.0-80k & 3.2       & 3.4   & 3.3 \\
M9-H-1.0-100k & 2.4      & 2.6   & 2.5 \\
M9-H-5.0-60k & 7.4       & 8.1   & 7.7 \\
M9-H-5.0-80k & 5.8       & 6.3   & 6.0 \\
M9-H-5.0-100k & 4.9      & 5.3   & 5.0 \\
M9-H-10.0-60k & 13.9     & 15.1  & 14.5 \\
M9-H-10.0-80k & 11.5     & 12.5  & 12.0 \\
M9-H-10.0-100k & 10.1    & 10.9  & 10.4 \\
M9-K-1.0-60k & 5.5       & 6.2   & 5.9 \\
M9-K-1.0-80k & 3.7       & 4.2   & 4.0 \\
M9-K-1.0-100k & 2.8      & 3.2   & 3.0 \\
M9-K-5.0-60k & 9.0       & 10.1  & 9.6 \\
M9-K-5.0-80k & 6.9       & 7.8   & 7.4 \\
M9-K-5.0-100k & 5.9      & 6.6   & 6.2 \\
M9-K-10.0-60k & 17.1     & 19.3  & 18.3 \\
M9-K-10.0-80k & 14.2     & 16.0  & 15.1 \\
M9-K-10.0-100k & 12.4    & 14.0  & 13.2 \\

\end{longtable}

\end{appendix}

\begin{appendix}

\section{Model spectra}\label{AppendixPlot}

\begin{figure*}

\includegraphics[width=16cm]{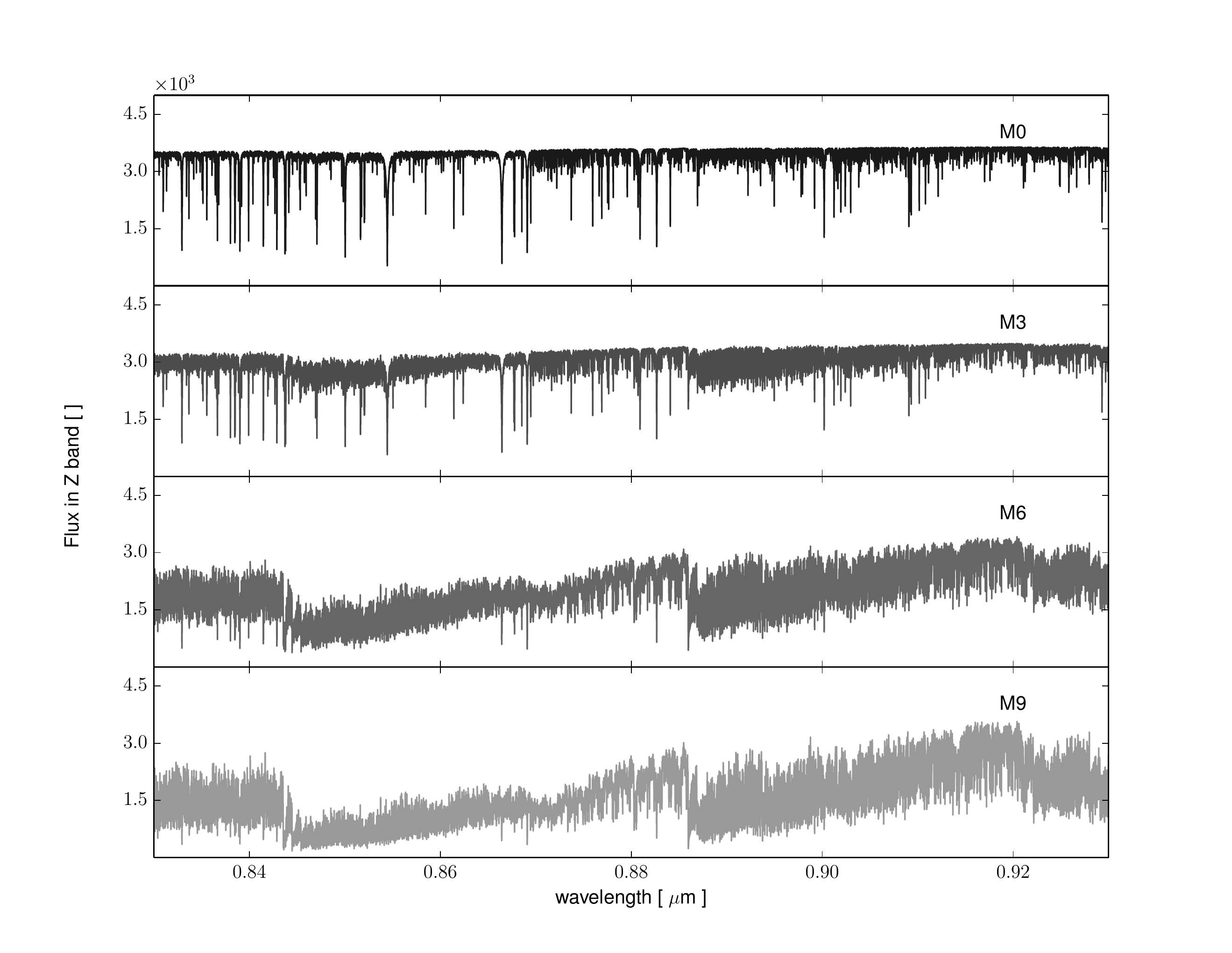}

\caption{Flux as a function of wavelength in the Z band for the $v.sin{i}$\,=\,1.0\,km/s and spectral types M0, M3, M6, and M9 ({\it top} to {\it bottom} panels) spectra, when seen at a resolution at 100\,000. Flux units are arbitrary.}\label{Zband_flux}

\end{figure*}

\begin{figure*}

\includegraphics[width=16cm]{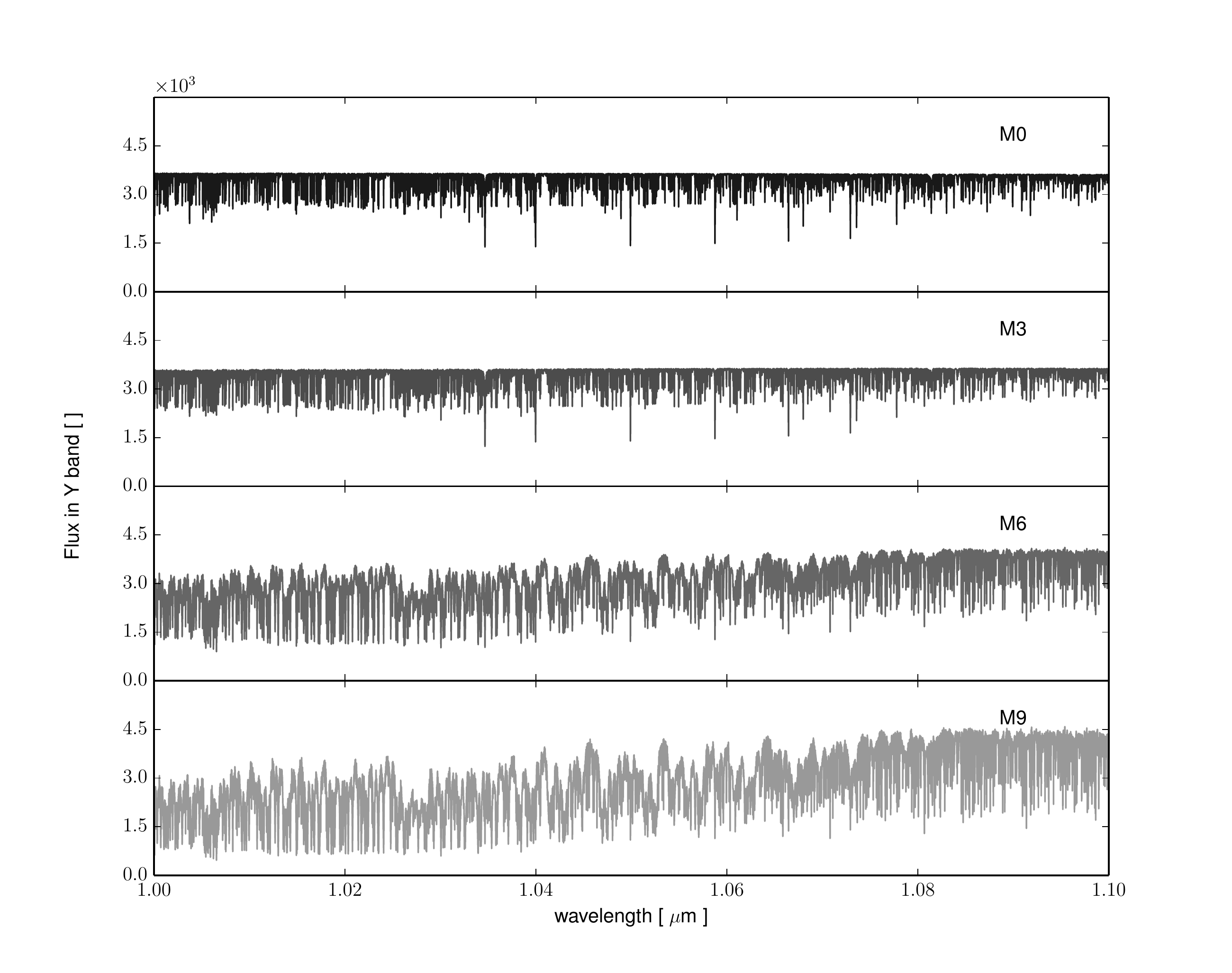}

\caption{Flux as a function of wavelength in the Y band for the $v.sin{i}$\,=\,1.0\,km/s and spectral types M0, M3, M6, and M9 ({\it top} to {\it bottom} panels) spectra, when seen at a resolution at 100\,000. Flux units are arbitrary.}\label{Yband_flux}

\end{figure*}

\begin{figure*}

\includegraphics[width=16cm]{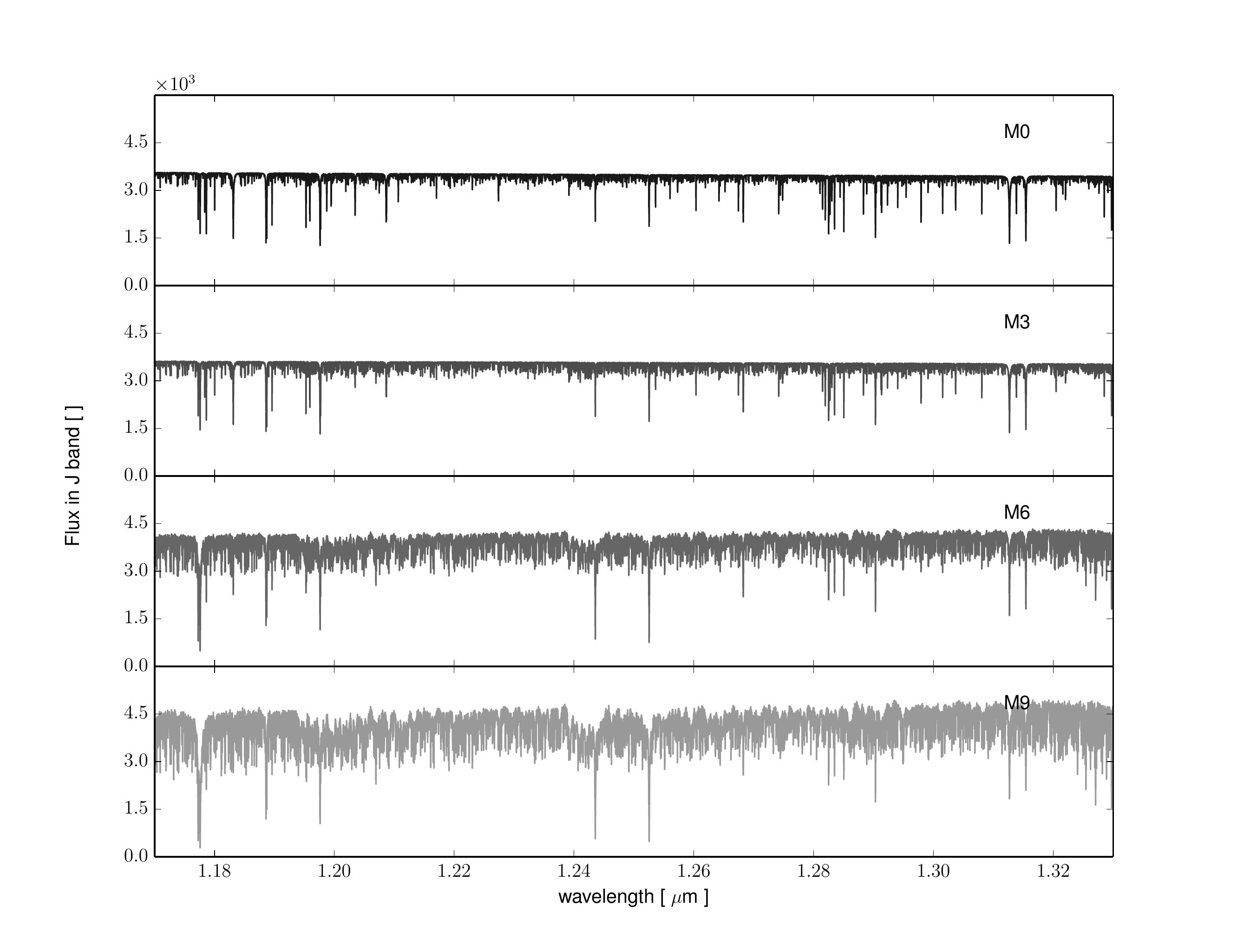}

\caption{Flux as a function of wavelength in the J band for the $v.sin{i}$\,=\,1.0\,km/s and spectral types M0, M3, M6, and M9 ({\it top} to {\it bottom} panels) spectra, when seen at a resolution at 100\,000. Flux units are arbitrary.}\label{Jband_flux}

\end{figure*}

\begin{figure*}

\includegraphics[width=16cm]{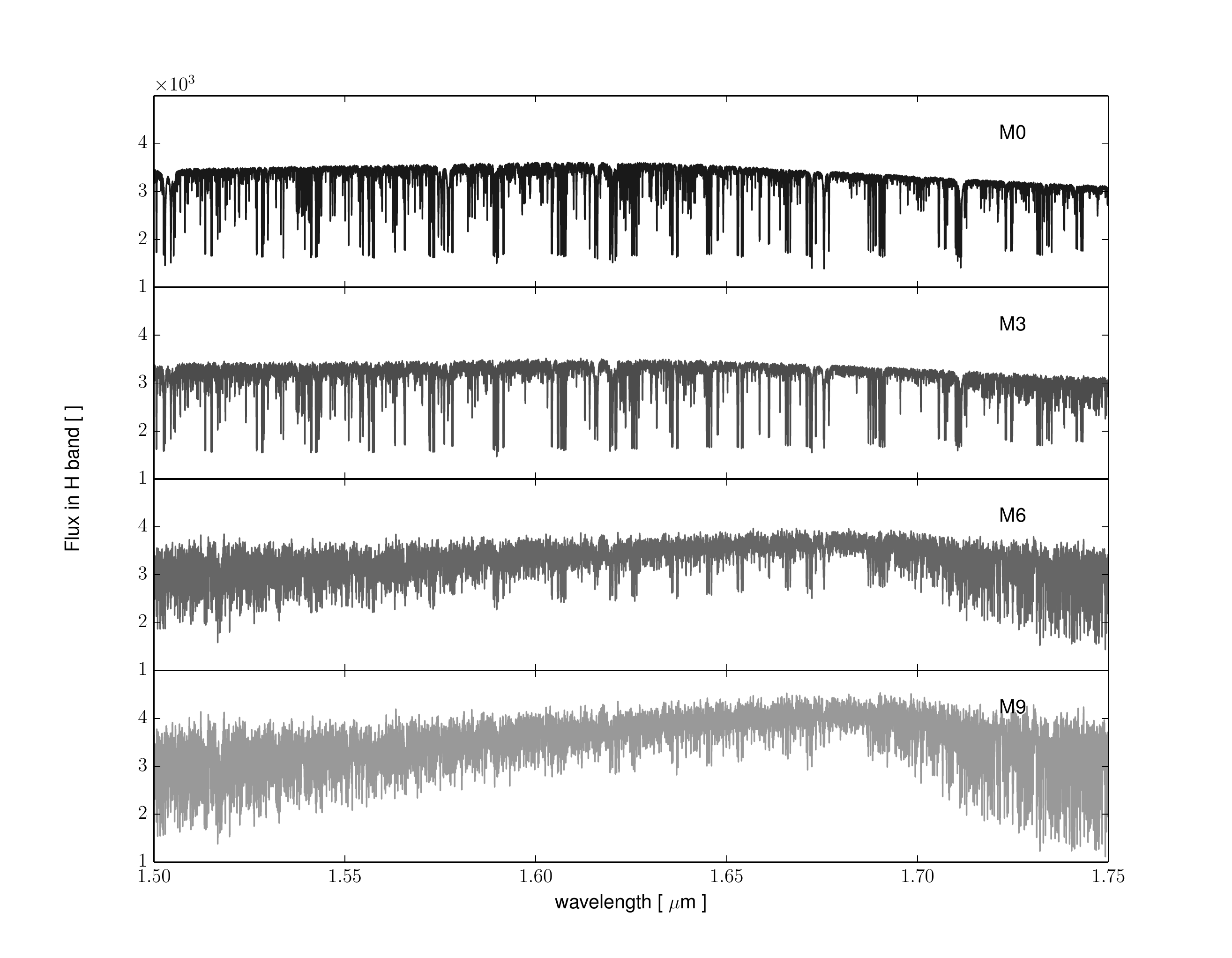}

\caption{Flux as a function of wavelength in the H band for the $v.sin{i}$\,=\,1.0\,km/s and spectral types M0, M3, M6, and M9 ({\it top} to {\it bottom} panels) spectra, when seen at a resolution at 100\,000. Flux units are arbitrary.}\label{Hband_flux}

\end{figure*}

\begin{figure*}

\includegraphics[width=16cm]{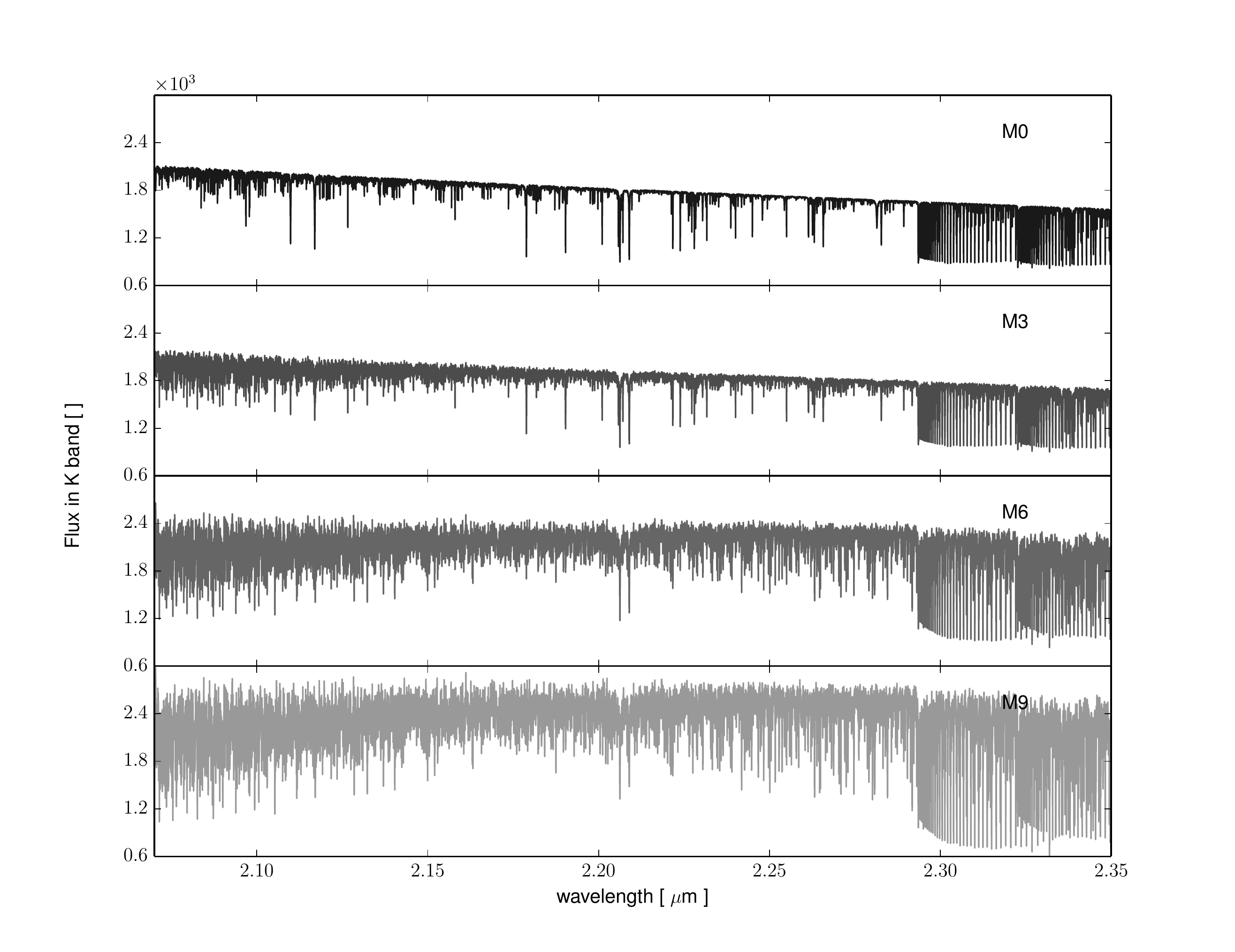}

\caption{Flux as a function of wavelength in the K band for the $v.sin{i}$\,=\,1.0\,km/s and spectral types M0, M3, M6, and M9 ({\it top} to {\it bottom} panels) spectra, when seen at a resolution at 100\,000. Flux units are arbitrary.}\label{Kband_flux}

\end{figure*}

In this section we plot examples of the PHOENIX-ACES models considered for a $v.sin{i}$\,=\,1.0\,km/s and a resolution of 100\,000 and for the different photometric bands considered for our work: ZYJHK in Fig.\,\ref{Zband_flux}\,\ref{Yband_flux}\,\ref{Jband_flux}\,\ref{Hband_flux},\, and \ref{Kband_flux}. We also plot in Fig. A.6 the atmospheric transmission of Earth's atmosphere as a function of wavelength as computed by the models considered in this paper for each of the bands considered. For a detailed identification of the most important features, we refer to Fig.\,1 of \cite{2015A&A...576A..77S}.

\begin{figure*}

\includegraphics[width=18cm]{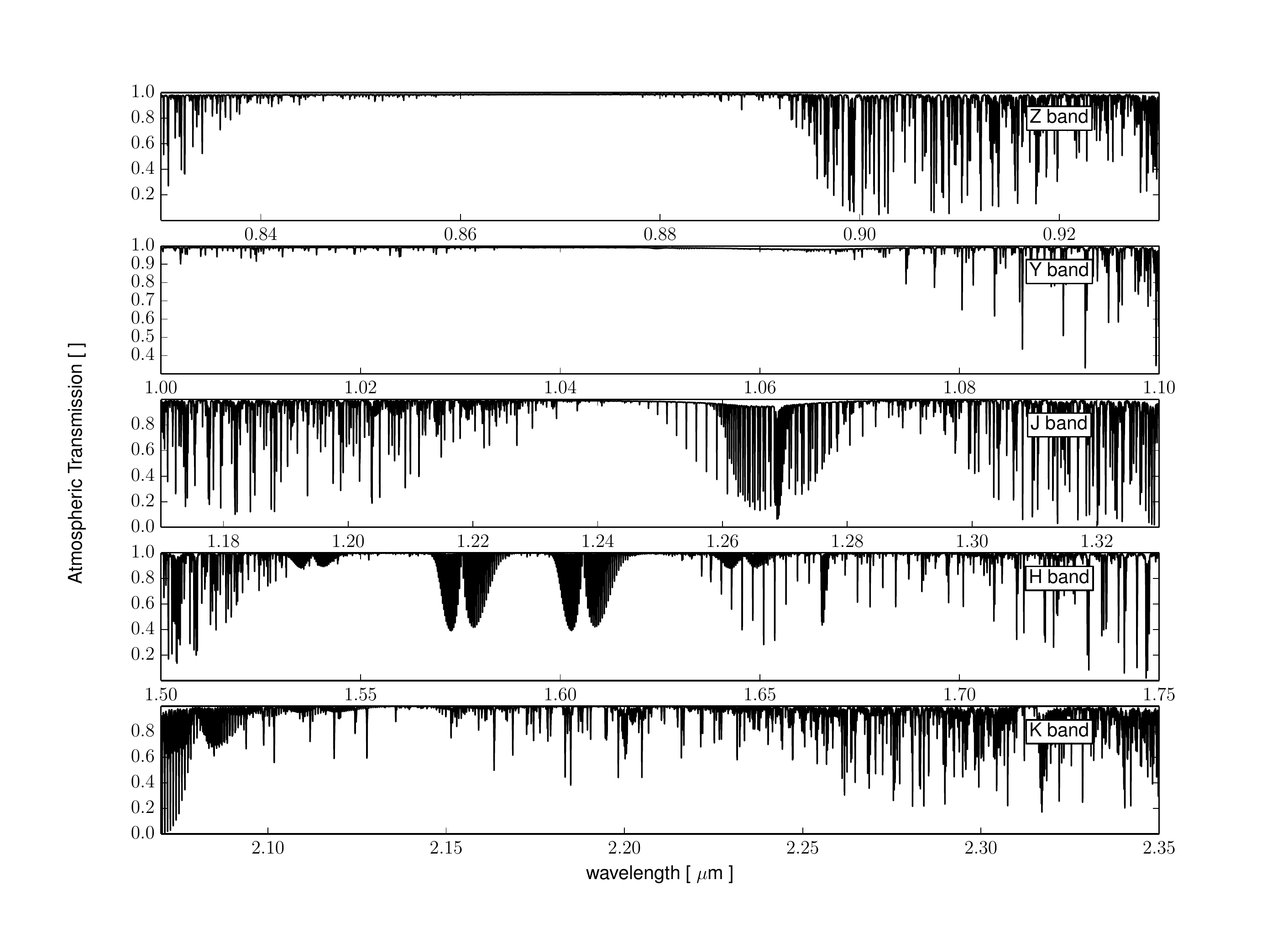}

\caption{Telluric transmission spectra as function of wavelength for each of the photometric bands considered in this work (ZYJHK, {\it top to bottom}).}\label{AT}

\end{figure*}

\end{appendix}

\begin{appendix}

\section{Different parameters on the aplication of condition 2}

\begin{figure*}
\centering
\includegraphics[width=16cm]{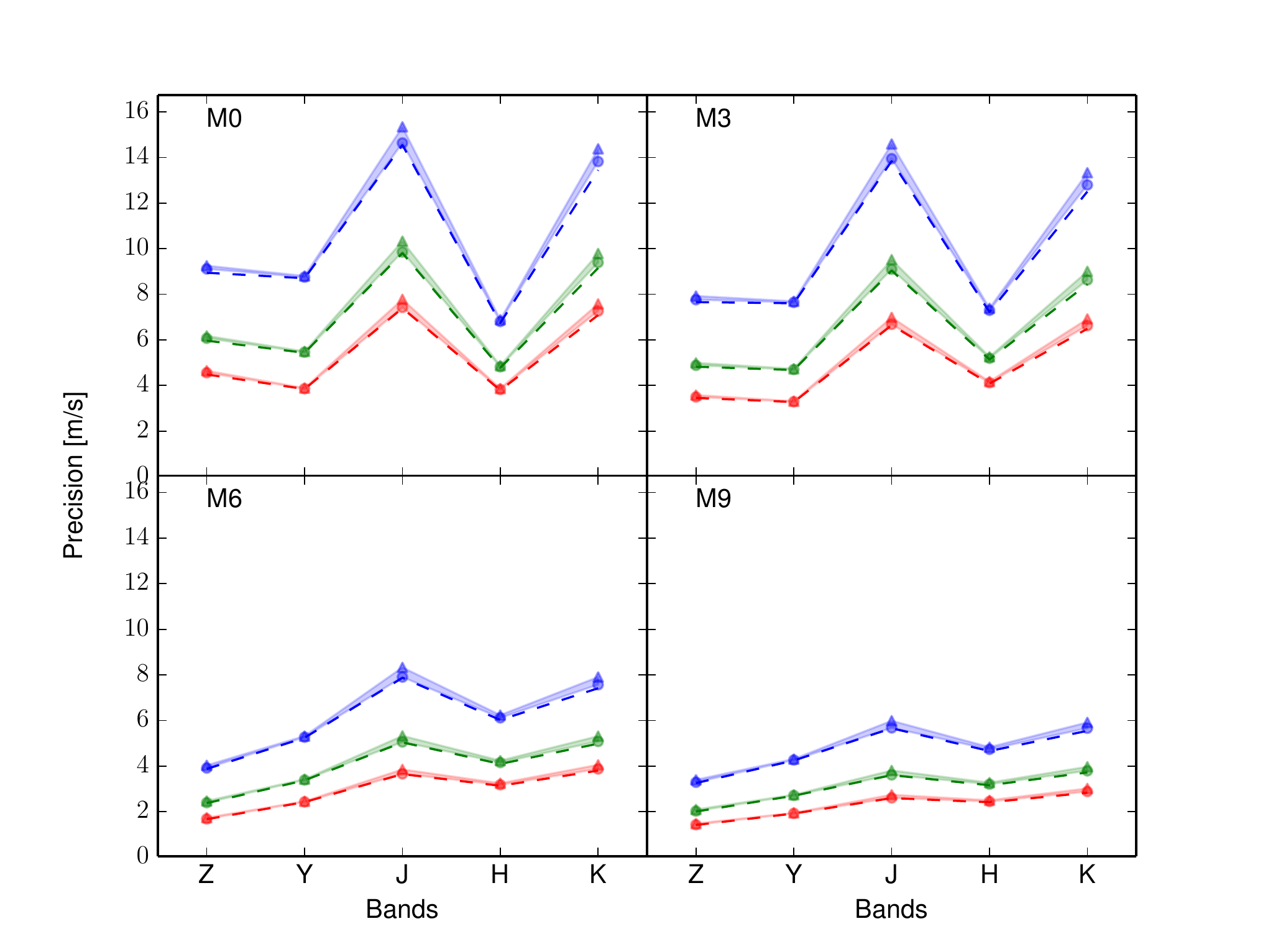}

\caption{Precision achieved as a function of spectral band for stars with a  rotational velocity of $v.\sin{i}$\,=\,1.0\,km/s and spectral types M0, M3, M6, and M9. The dashed line represents the theoretical limits imposed by condition 1 and the filled area represents the values within the limits set by conditions 2 ({\it circles}) and 3 ({\it triangles}); blue, green and red represent the results obtained for resolutions of 60\,000, 80\,000, and 100\,000, respectively. The spectra were normalized to have a S/N of 100 per resolution element as measured at the center of the J band (see Sect.\,\ref{simuls} for details). For the application of condition 2, the wavelengths with four consecutive atmospheric transmission pixels within 30\,km/s were considered.}\label{Res_4pixel}

\end{figure*}

\begin{figure*}
\centering
\includegraphics[width=16cm]{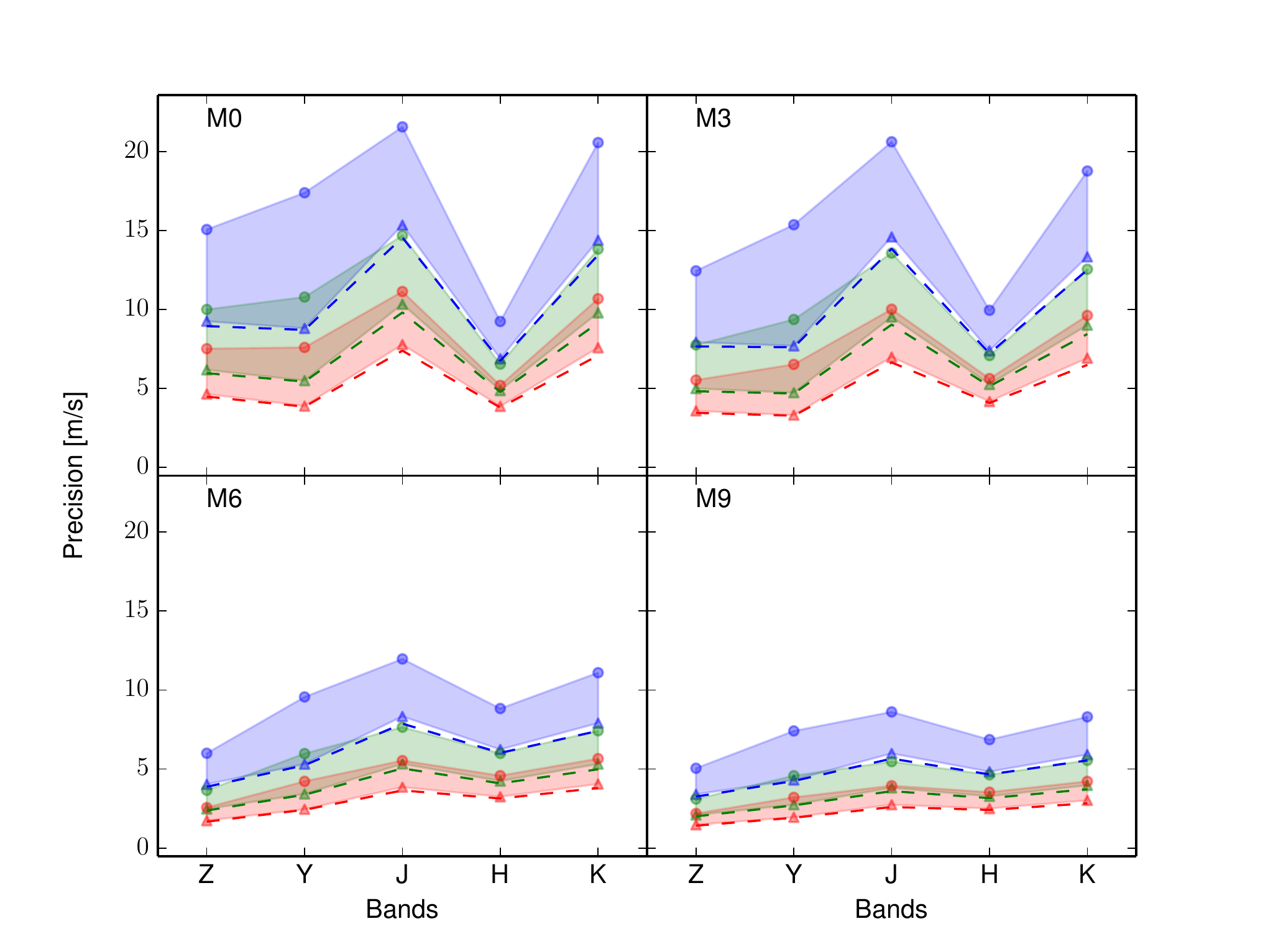}

\caption{Precision achieved as a function of spectral band for stars with a  rotational velocity of $v.\sin{i}$\,=\,1.0\,km/s and spectral types M0, M3, M6, and M9. The dashed line represents the theoretical limits imposed by condition 1 and the filled area represents the values within the limits set by condition 2 ({\it circles}) and 3 ({\it triangles}); blue, green and red represent the results obtained for resolutions of 60\,000, 80\,000, and 100\,000, respectively. The spectra were normalized to have a S/N of 100 per resolution element as measured at the center of the J band (see Sect.\,\ref{simuls} for details). For the application of condition 2, the wavelengths with two consecutive atmospheric transmission pixels within 30\,km/s were considered.}\label{Res_2pixel}

\end{figure*}

\begin{figure*}
\centering
\includegraphics[width=16cm]{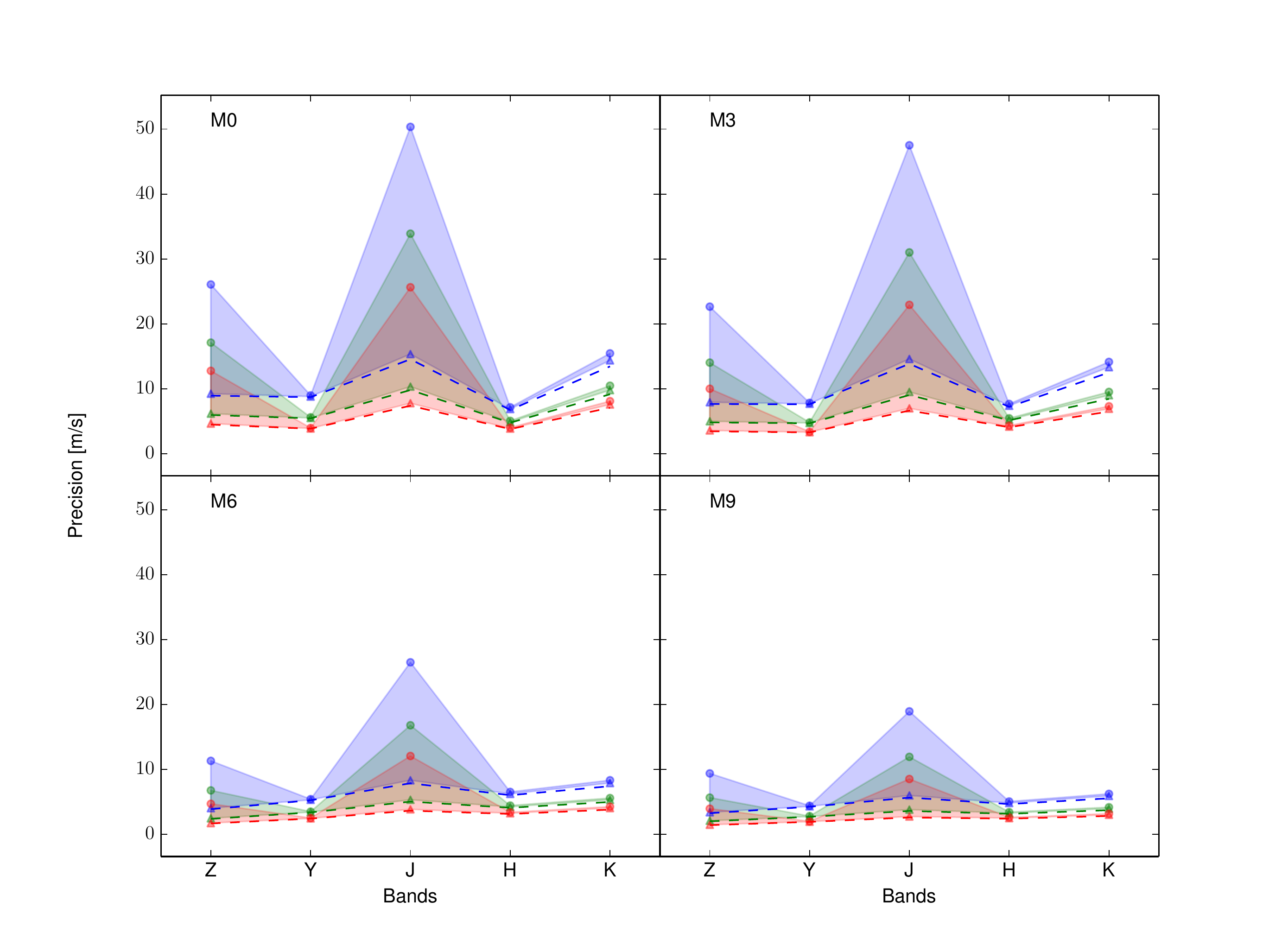}

\caption{Precision achieved as a function of spectral band for stars with a  rotational velocity of $v.\sin{i}$\,=\,1.0\,km/s and spectral types M0, M3, M6, and M9. The dashed line represents the theoretical limits imposed by condition 1 and the filled area represents the values within the limits set by condition 2 ({\it circles}) and 3 ({\it triangles}); blue, green and red represent the results obtained for resolutions of 60\,000, 80\,000, and 100\,000, respectively. The spectra were normalized to have a S/N of 100 per resolution element as measured at the center of the J band (see Sect.\,\ref{simuls} for details). For the application of condition 2, the spectra were considered to be shifted by -\,10\,km/s.}\label{Res_vel-10}

\end{figure*}

\begin{figure*}
\centering
\includegraphics[width=16cm]{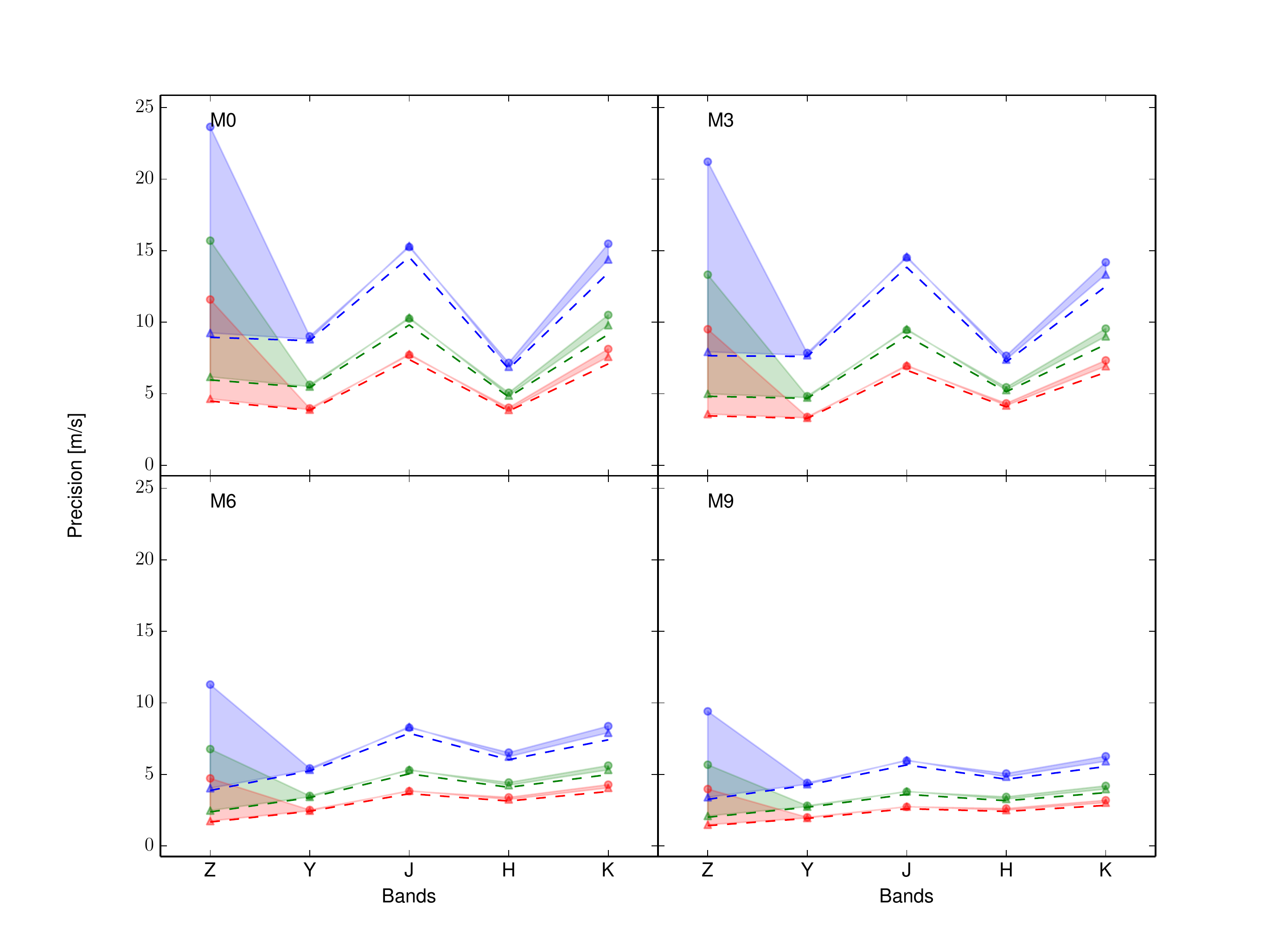}

\caption{Precision achieved as a function of spectral band for stars with a  rotational velocity of $v.\sin{i}$\,=\,1.0\,km/s and spectral types M0, M3, M6, and M9. The dashed line represents the theoretical limits imposed by condition 1 and the filled area represents the values within the limits set by condition 2 ({\it circles}) and 3 ({\it triangles}); blue, green and red represent the results obtained for resolutions of 60\,000, 80\,000, and 100\,000, respectively. The spectra were normalized to have a S/N of 100 per resolution element as measured at the center of the J band (see Sect.\,\ref{simuls} for details). For the application of condition 2, the spectra were considered to be shifted by +\,10\,km/s.}\label{Res_vel+10}

\end{figure*}

In this section we present the result of alternative runs of our RV precision calculation in which we considered
\begin{itemize}
 \item a different number of consecutive pixels required for the flagging of a wavelength as being below the threshold and thus discard it for the calculation of RV precision following condition 2: four pixels in Fig.\,\ref{Res_4pixel} and two pixels in Fig.\,\ref{Res_2pixel}\,;
 \item different systemic velocities of the star (for three consecutive pixels): -10\,km/s in Fig.\,\ref{Res_vel-10} and +10\,km/s in in Fig.\,\ref{Res_vel+10}\,.
\end{itemize}

These cases lead to different achievable precision as reported by condition 2, showing that it is very vulnerable to the particular parametrization of the problem.

\end{appendix}

\end{document}